\documentclass[conference]{IEEEtran}
\IEEEoverridecommandlockouts

\newif\ifextended
\extendedtrue

\newif\ifdraft
\draftfalse

\newif\ifblind
\blindfalse

\ifextended
\newcommand{\titledpar}[1]{\addvspace{5pt}\noindent\textbf{#1}}
\newcommand{\titledsubpar}[1]{\addvspace{5pt}\noindent\textit{#1}}
\else
  \newcommand{\titledpar}[1]{\addvspace{2pt}\noindent\textbf{#1}}
  \newcommand{\titledsubpar}[1]{\addvspace{1pt}\noindent\textit{#1}}
\fi

\usepackage[T1]{fontenc}
\usepackage[utf8]{inputenc}
\usepackage[scaled=0.81]{beramono}  

\usepackage{hyperref}
\usepackage{cite}
\usepackage{booktabs,multirow,multicol,array}
\usepackage{relsize}
\usepackage[skip=1pt]{subcaption}  \usepackage{graphicx,xcolor}
\usepackage{numprint}
\npdecimalsign{.}
\usepackage{amsmath,amssymb,amsfonts}
\usepackage{xurl}
\usepackage{etoolbox}
\makeatletter
\patchcmd{\@startsection}
  {\@afterindenttrue}  {\@afterindentfalse} {}{} \makeatother

\graphicspath{{./floats/}}
\DeclareGraphicsExtensions{.pdf,.png}

\usepackage{xparse,xstring,xspace}
\usepackage{dcolumn}
\usepackage{caption}

\usepackage[inline]{enumitem}
\setlist[enumerate]{label=\emph{\roman*})}
\setlist[description]{font=\normalfont\bfseries}

\usepackage{tikz}
\usetikzlibrary{shapes,arrows,positioning,calc,fit,intersections,through}

\DeclareDocumentCommand{\plotbar}{O{black} O{1ex} O{3pt}}{\tikz[baseline,yshift=#3] {
    \draw[#1, line width=2pt] (0, #2) -- (0, -#2);
  }\xspace }

\ifdraft
\usepackage[colorinlistoftodos,textsize=scriptsize,obeyFinal,textwidth=12mm]{todonotes}

\DeclareDocumentCommand{\ReviewNote}{s o m O{white}}{\todo[color=#4,\IfBooleanTF{#1}{inline}{}]{\IfNoValueF{#2}{\textbf{#2:}\xspace}#3}
}
\else
\DeclareDocumentCommand{\ReviewNote}{s o m O{white}}{}
\fi

\DeclareDocumentCommand{\caf}{s m}{\IfBooleanTF{#1}{\ReviewNote*{#2}[yellow]}{\ReviewNote{#2}[yellow]}}
\DeclareDocumentCommand{\am}{s m}{\IfBooleanTF{#1}{\ReviewNote*{#2}[red!70!white]}{\ReviewNote{#2}[red!70!white]}}

\definecolor{plays}{HTML}{1B9E77}
\definecolor{timing}{HTML}{D95F02}
\definecolor{style}{HTML}{7570B3}
\definecolor{difficulty}{HTML}{E7298A}
\definecolor{user}{HTML}{66A61E}
\definecolor{negative}{HTML}{7b3294}
\definecolor{positive}{HTML}{008837}
 \definecolor{mlcol}{HTML}{e41a1c}
\definecolor{mucol}{HTML}{377eb8}

\DeclareDocumentCommand{\vg}{s m o}{\IfBooleanTF{#1}{\var{\textcolor{#2}{\IfNoValueTF{#3}{#2}{#3}}}}{\textcolor{#2}{\IfNoValueTF{#3}{#2}{#3}}}}

\usepackage{pgfkeys,pgfplots,numprint,relsize}
\xspaceaddexceptions{\%}  \def\NAN{??}              \def\keyfamily{/sea/}     

\usepackage{fmtcount}     

\pgfkeyssetvalue{/sea/N.raw.vars}{36}
\pgfkeyssetvalue{/sea/N.all.variables}{22}
\pgfkeyssetvalue{/sea/N.levels}{26609725}
\pgfkeyssetvalue{/sea/N.fml.all.variables}{13}
\pgfkeyssetvalue{/sea/N.fml.all.clears.variables}{14}
\pgfkeyssetvalue{/sea/N.fml.all.attempts.variables}{14}
\pgfkeyssetvalue{/sea/N.fml.all.both.variables}{15}
\pgfkeyssetvalue{/sea/N.all.user.vars}{19}
\pgfkeyssetvalue{/sea/N.users}{5935222}
\pgfkeyssetvalue{/sea/N.all.makers}{2635933}
\pgfkeyssetvalue{/sea/N.fml.user.all.variables}{28}
\pgfkeyssetvalue{/sea/N.user.vars}{28}
\pgfkeyssetvalue{/sea/aic:user/fml.all.both}{-31284103.9032247}
\pgfkeyssetvalue{/sea/aic:user/fml.user.all}{-31664107.3053527}
\pgfkeyssetvalue{/sea/resdev:user/fml.all.both}{480817.654006252}
\pgfkeyssetvalue{/sea/resdev:user/fml.user.all}{473999.540342909}
\pgfkeyssetvalue{/sea/numeric/min:world_record}{-0.001}
\pgfkeyssetvalue{/sea/numeric/min:upload_time}{0.015}
\pgfkeyssetvalue{/sea/numeric/min:upload_attempts}{1}
\pgfkeyssetvalue{/sea/numeric/min:num_comments}{0}
\pgfkeyssetvalue{/sea/numeric/min:timer}{10}
\pgfkeyssetvalue{/sea/numeric/min:autoscroll_speed}{0}
\pgfkeyssetvalue{/sea/numeric/min:clears}{0}
\pgfkeyssetvalue{/sea/numeric/min:attempts}{0}
\pgfkeyssetvalue{/sea/numeric/min:clear_rate_p}{0}
\pgfkeyssetvalue{/sea/numeric/min:likes}{0}
\pgfkeyssetvalue{/sea/numeric/min:boos}{0}
\pgfkeyssetvalue{/sea/numeric/min:unique_players_and_versus}{0}
\pgfkeyssetvalue{/sea/numeric/min:courses_played}{0}
\pgfkeyssetvalue{/sea/numeric/min:courses_cleared}{0}
\pgfkeyssetvalue{/sea/numeric/min:courses_attempted}{0}
\pgfkeyssetvalue{/sea/numeric/min:courses_deaths}{0}
\pgfkeyssetvalue{/sea/numeric/min:likes_maker}{0}
\pgfkeyssetvalue{/sea/numeric/min:maker_points}{0}
\pgfkeyssetvalue{/sea/numeric/min:easy_highscore}{0}
\pgfkeyssetvalue{/sea/numeric/min:normal_highscore}{0}
\pgfkeyssetvalue{/sea/numeric/min:expert_highscore}{0}
\pgfkeyssetvalue{/sea/numeric/min:super_expert_highscore}{0}
\pgfkeyssetvalue{/sea/numeric/min:versus_rating}{0}
\pgfkeyssetvalue{/sea/numeric/min:first_clears}{0}
\pgfkeyssetvalue{/sea/numeric/min:world_records}{0}
\pgfkeyssetvalue{/sea/numeric/min:uploaded_levels}{0}
\pgfkeyssetvalue{/sea/numeric/25p:world_record}{9.836}
\pgfkeyssetvalue{/sea/numeric/25p:upload_time}{21.899}
\pgfkeyssetvalue{/sea/numeric/25p:upload_attempts}{1}
\pgfkeyssetvalue{/sea/numeric/25p:num_comments}{0}
\pgfkeyssetvalue{/sea/numeric/25p:timer}{300}
\pgfkeyssetvalue{/sea/numeric/25p:autoscroll_speed}{0}
\pgfkeyssetvalue{/sea/numeric/25p:clears}{4}
\pgfkeyssetvalue{/sea/numeric/25p:attempts}{22}
\pgfkeyssetvalue{/sea/numeric/25p:clear_rate_p}{0.0958466453674121}
\pgfkeyssetvalue{/sea/numeric/25p:likes}{0}
\pgfkeyssetvalue{/sea/numeric/25p:boos}{0}
\pgfkeyssetvalue{/sea/numeric/25p:unique_players_and_versus}{10}
\pgfkeyssetvalue{/sea/numeric/25p:courses_played}{127}
\pgfkeyssetvalue{/sea/numeric/25p:courses_cleared}{80}
\pgfkeyssetvalue{/sea/numeric/25p:courses_attempted}{508}
\pgfkeyssetvalue{/sea/numeric/25p:courses_deaths}{376}
\pgfkeyssetvalue{/sea/numeric/25p:likes_maker}{23}
\pgfkeyssetvalue{/sea/numeric/25p:maker_points}{780}
\pgfkeyssetvalue{/sea/numeric/25p:easy_highscore}{6}
\pgfkeyssetvalue{/sea/numeric/25p:normal_highscore}{0}
\pgfkeyssetvalue{/sea/numeric/25p:expert_highscore}{0}
\pgfkeyssetvalue{/sea/numeric/25p:super_expert_highscore}{0}
\pgfkeyssetvalue{/sea/numeric/25p:versus_rating}{0}
\pgfkeyssetvalue{/sea/numeric/25p:first_clears}{1}
\pgfkeyssetvalue{/sea/numeric/25p:world_records}{0}
\pgfkeyssetvalue{/sea/numeric/25p:uploaded_levels}{10}
\pgfkeyssetvalue{/sea/numeric/50p:world_record}{21.032}
\pgfkeyssetvalue{/sea/numeric/50p:upload_time}{42.361}
\pgfkeyssetvalue{/sea/numeric/50p:upload_attempts}{1}
\pgfkeyssetvalue{/sea/numeric/50p:num_comments}{0}
\pgfkeyssetvalue{/sea/numeric/50p:timer}{300}
\pgfkeyssetvalue{/sea/numeric/50p:autoscroll_speed}{0}
\pgfkeyssetvalue{/sea/numeric/50p:clears}{10}
\pgfkeyssetvalue{/sea/numeric/50p:attempts}{54}
\pgfkeyssetvalue{/sea/numeric/50p:clear_rate_p}{0.25}
\pgfkeyssetvalue{/sea/numeric/50p:likes}{1}
\pgfkeyssetvalue{/sea/numeric/50p:boos}{1}
\pgfkeyssetvalue{/sea/numeric/50p:unique_players_and_versus}{21}
\pgfkeyssetvalue{/sea/numeric/50p:courses_played}{342}
\pgfkeyssetvalue{/sea/numeric/50p:courses_cleared}{219}
\pgfkeyssetvalue{/sea/numeric/50p:courses_attempted}{1397}
\pgfkeyssetvalue{/sea/numeric/50p:courses_deaths}{1064}
\pgfkeyssetvalue{/sea/numeric/50p:likes_maker}{71}
\pgfkeyssetvalue{/sea/numeric/50p:maker_points}{1599}
\pgfkeyssetvalue{/sea/numeric/50p:easy_highscore}{29}
\pgfkeyssetvalue{/sea/numeric/50p:normal_highscore}{0}
\pgfkeyssetvalue{/sea/numeric/50p:expert_highscore}{0}
\pgfkeyssetvalue{/sea/numeric/50p:super_expert_highscore}{0}
\pgfkeyssetvalue{/sea/numeric/50p:versus_rating}{308}
\pgfkeyssetvalue{/sea/numeric/50p:first_clears}{4}
\pgfkeyssetvalue{/sea/numeric/50p:world_records}{2}
\pgfkeyssetvalue{/sea/numeric/50p:uploaded_levels}{25}
\pgfkeyssetvalue{/sea/numeric/75p:world_record}{43.85}
\pgfkeyssetvalue{/sea/numeric/75p:upload_time}{77.423}
\pgfkeyssetvalue{/sea/numeric/75p:upload_attempts}{4}
\pgfkeyssetvalue{/sea/numeric/75p:num_comments}{1}
\pgfkeyssetvalue{/sea/numeric/75p:timer}{300}
\pgfkeyssetvalue{/sea/numeric/75p:autoscroll_speed}{0}
\pgfkeyssetvalue{/sea/numeric/75p:clears}{28}
\pgfkeyssetvalue{/sea/numeric/75p:attempts}{127}
\pgfkeyssetvalue{/sea/numeric/75p:clear_rate_p}{0.5}
\pgfkeyssetvalue{/sea/numeric/75p:likes}{4}
\pgfkeyssetvalue{/sea/numeric/75p:boos}{2}
\pgfkeyssetvalue{/sea/numeric/75p:unique_players_and_versus}{48}
\pgfkeyssetvalue{/sea/numeric/75p:courses_played}{840}
\pgfkeyssetvalue{/sea/numeric/75p:courses_cleared}{534}
\pgfkeyssetvalue{/sea/numeric/75p:courses_attempted}{3391}
\pgfkeyssetvalue{/sea/numeric/75p:courses_deaths}{2654}
\pgfkeyssetvalue{/sea/numeric/75p:likes_maker}{174}
\pgfkeyssetvalue{/sea/numeric/75p:maker_points}{2441}
\pgfkeyssetvalue{/sea/numeric/75p:easy_highscore}{100}
\pgfkeyssetvalue{/sea/numeric/75p:normal_highscore}{4}
\pgfkeyssetvalue{/sea/numeric/75p:expert_highscore}{1}
\pgfkeyssetvalue{/sea/numeric/75p:super_expert_highscore}{0}
\pgfkeyssetvalue{/sea/numeric/75p:versus_rating}{1865}
\pgfkeyssetvalue{/sea/numeric/75p:first_clears}{11}
\pgfkeyssetvalue{/sea/numeric/75p:world_records}{7}
\pgfkeyssetvalue{/sea/numeric/75p:uploaded_levels}{52}
\pgfkeyssetvalue{/sea/numeric/99p:world_record}{258.211759999998}
\pgfkeyssetvalue{/sea/numeric/99p:upload_time}{261.384}
\pgfkeyssetvalue{/sea/numeric/99p:upload_attempts}{45}
\pgfkeyssetvalue{/sea/numeric/99p:num_comments}{13}
\pgfkeyssetvalue{/sea/numeric/99p:timer}{500}
\pgfkeyssetvalue{/sea/numeric/99p:autoscroll_speed}{0}
\pgfkeyssetvalue{/sea/numeric/99p:clears}{247}
\pgfkeyssetvalue{/sea/numeric/99p:attempts}{1431}
\pgfkeyssetvalue{/sea/numeric/99p:clear_rate_p}{1}
\pgfkeyssetvalue{/sea/numeric/99p:likes}{49}
\pgfkeyssetvalue{/sea/numeric/99p:boos}{15}
\pgfkeyssetvalue{/sea/numeric/99p:unique_players_and_versus}{325}
\pgfkeyssetvalue{/sea/numeric/99p:courses_played}{6844}
\pgfkeyssetvalue{/sea/numeric/99p:courses_cleared}{4501.75999999791}
\pgfkeyssetvalue{/sea/numeric/99p:courses_attempted}{34643}
\pgfkeyssetvalue{/sea/numeric/99p:courses_deaths}{29540}
\pgfkeyssetvalue{/sea/numeric/99p:likes_maker}{1896}
\pgfkeyssetvalue{/sea/numeric/99p:maker_points}{7687}
\pgfkeyssetvalue{/sea/numeric/99p:easy_highscore}{1408}
\pgfkeyssetvalue{/sea/numeric/99p:normal_highscore}{547}
\pgfkeyssetvalue{/sea/numeric/99p:expert_highscore}{23}
\pgfkeyssetvalue{/sea/numeric/99p:super_expert_highscore}{4}
\pgfkeyssetvalue{/sea/numeric/99p:versus_rating}{4715}
\pgfkeyssetvalue{/sea/numeric/99p:first_clears}{215}
\pgfkeyssetvalue{/sea/numeric/99p:world_records}{268}
\pgfkeyssetvalue{/sea/numeric/99p:uploaded_levels}{100}
\pgfkeyssetvalue{/sea/numeric/99.99p:world_record}{2627.83619320001}
\pgfkeyssetvalue{/sea/numeric/99.99p:upload_time}{494.049}
\pgfkeyssetvalue{/sea/numeric/99.99p:upload_attempts}{378.027600001544}
\pgfkeyssetvalue{/sea/numeric/99.99p:num_comments}{906.055200003088}
\pgfkeyssetvalue{/sea/numeric/99.99p:timer}{500}
\pgfkeyssetvalue{/sea/numeric/99.99p:autoscroll_speed}{2}
\pgfkeyssetvalue{/sea/numeric/99.99p:clears}{15557.2208000124}
\pgfkeyssetvalue{/sea/numeric/99.99p:attempts}{142162.7940001}
\pgfkeyssetvalue{/sea/numeric/99.99p:clear_rate_p}{1}
\pgfkeyssetvalue{/sea/numeric/99.99p:likes}{4590.02760000154}
\pgfkeyssetvalue{/sea/numeric/99.99p:boos}{576}
\pgfkeyssetvalue{/sea/numeric/99.99p:unique_players_and_versus}{18171.1104000062}
\pgfkeyssetvalue{/sea/numeric/99.99p:courses_played}{63356}
\pgfkeyssetvalue{/sea/numeric/99.99p:courses_cleared}{57813}
\pgfkeyssetvalue{/sea/numeric/99.99p:courses_attempted}{452109}
\pgfkeyssetvalue{/sea/numeric/99.99p:courses_deaths}{422060}
\pgfkeyssetvalue{/sea/numeric/99.99p:likes_maker}{241521}
\pgfkeyssetvalue{/sea/numeric/99.99p:maker_points}{22633}
\pgfkeyssetvalue{/sea/numeric/99.99p:easy_highscore}{30264}
\pgfkeyssetvalue{/sea/numeric/99.99p:normal_highscore}{16501}
\pgfkeyssetvalue{/sea/numeric/99.99p:expert_highscore}{5700}
\pgfkeyssetvalue{/sea/numeric/99.99p:super_expert_highscore}{291}
\pgfkeyssetvalue{/sea/numeric/99.99p:versus_rating}{7004}
\pgfkeyssetvalue{/sea/numeric/99.99p:first_clears}{7803}
\pgfkeyssetvalue{/sea/numeric/99.99p:world_records}{10674}
\pgfkeyssetvalue{/sea/numeric/99.99p:uploaded_levels}{100}
\pgfkeyssetvalue{/sea/numeric/max:world_record}{5999.999}
\pgfkeyssetvalue{/sea/numeric/max:upload_time}{499.987}
\pgfkeyssetvalue{/sea/numeric/max:upload_attempts}{5983}
\pgfkeyssetvalue{/sea/numeric/max:num_comments}{120515}
\pgfkeyssetvalue{/sea/numeric/max:timer}{500}
\pgfkeyssetvalue{/sea/numeric/max:autoscroll_speed}{2}
\pgfkeyssetvalue{/sea/numeric/max:clears}{1804430}
\pgfkeyssetvalue{/sea/numeric/max:attempts}{67895905}
\pgfkeyssetvalue{/sea/numeric/max:clear_rate_p}{1}
\pgfkeyssetvalue{/sea/numeric/max:likes}{564890}
\pgfkeyssetvalue{/sea/numeric/max:boos}{37886}
\pgfkeyssetvalue{/sea/numeric/max:unique_players_and_versus}{1302862}
\pgfkeyssetvalue{/sea/numeric/max:courses_played}{549016}
\pgfkeyssetvalue{/sea/numeric/max:courses_cleared}{513124}
\pgfkeyssetvalue{/sea/numeric/max:courses_attempted}{17110430}
\pgfkeyssetvalue{/sea/numeric/max:courses_deaths}{17095589}
\pgfkeyssetvalue{/sea/numeric/max:likes_maker}{1278573}
\pgfkeyssetvalue{/sea/numeric/max:maker_points}{29100}
\pgfkeyssetvalue{/sea/numeric/max:easy_highscore}{320500}
\pgfkeyssetvalue{/sea/numeric/max:normal_highscore}{109304}
\pgfkeyssetvalue{/sea/numeric/max:expert_highscore}{77777}
\pgfkeyssetvalue{/sea/numeric/max:super_expert_highscore}{20000}
\pgfkeyssetvalue{/sea/numeric/max:versus_rating}{8000}
\pgfkeyssetvalue{/sea/numeric/max:first_clears}{67102}
\pgfkeyssetvalue{/sea/numeric/max:world_records}{105379}
\pgfkeyssetvalue{/sea/numeric/max:uploaded_levels}{100}
\pgfkeyssetvalue{/sea/factor/difficulty_nameEasy:Count}{8710121}
\pgfkeyssetvalue{/sea/factor/difficulty_nameEasy:Percentage}{32.7328486107992}
\pgfkeyssetvalue{/sea/factor/difficulty_nameNormal:Count}{11744359}
\pgfkeyssetvalue{/sea/factor/difficulty_nameNormal:Percentage}{44.1355895260098}
\pgfkeyssetvalue{/sea/factor/difficulty_nameExpert:Count}{4150458}
\pgfkeyssetvalue{/sea/factor/difficulty_nameExpert:Percentage}{15.5975230860146}
\pgfkeyssetvalue{/sea/factor/difficulty_nameSuper expert:Count}{2004787}
\pgfkeyssetvalue{/sea/factor/difficulty_nameSuper expert:Percentage}{7.53403877717639}
\pgfkeyssetvalue{/sea/factor/gamestyle_nameSMB1:Count}{3739380}
\pgfkeyssetvalue{/sea/factor/gamestyle_nameSMB1:Percentage}{14.0526818672497}
\pgfkeyssetvalue{/sea/factor/gamestyle_nameSMB3:Count}{2039702}
\pgfkeyssetvalue{/sea/factor/gamestyle_nameSMB3:Percentage}{7.66525020457746}
\pgfkeyssetvalue{/sea/factor/gamestyle_nameSMW:Count}{4572845}
\pgfkeyssetvalue{/sea/factor/gamestyle_nameSMW:Percentage}{17.1848638044925}
\pgfkeyssetvalue{/sea/factor/gamestyle_nameNSMBU:Count}{8750385}
\pgfkeyssetvalue{/sea/factor/gamestyle_nameNSMBU:Percentage}{32.8841617115547}
\pgfkeyssetvalue{/sea/factor/gamestyle_nameSM3DW:Count}{7507413}
\pgfkeyssetvalue{/sea/factor/gamestyle_nameSM3DW:Percentage}{28.2130424121256}
\pgfkeyssetvalue{/sea/factor/theme_nameAirship:Count}{1647068}
\pgfkeyssetvalue{/sea/factor/theme_nameAirship:Percentage}{6.18972199073835}
\pgfkeyssetvalue{/sea/factor/theme_nameCastle:Count}{3978403}
\pgfkeyssetvalue{/sea/factor/theme_nameCastle:Percentage}{14.9509361708924}
\pgfkeyssetvalue{/sea/factor/theme_nameDesert:Count}{1770880}
\pgfkeyssetvalue{/sea/factor/theme_nameDesert:Percentage}{6.65501052716629}
\pgfkeyssetvalue{/sea/factor/theme_nameForest:Count}{2188222}
\pgfkeyssetvalue{/sea/factor/theme_nameForest:Percentage}{8.2233920117551}
\pgfkeyssetvalue{/sea/factor/theme_nameGhost house:Count}{1357902}
\pgfkeyssetvalue{/sea/factor/theme_nameGhost house:Percentage}{5.1030290617434}
\pgfkeyssetvalue{/sea/factor/theme_nameOverworld:Count}{8352189}
\pgfkeyssetvalue{/sea/factor/theme_nameOverworld:Percentage}{31.3877313651306}
\pgfkeyssetvalue{/sea/factor/theme_nameSky:Count}{2610614}
\pgfkeyssetvalue{/sea/factor/theme_nameSky:Percentage}{9.81075152035581}
\pgfkeyssetvalue{/sea/factor/theme_nameSnow:Count}{2317182}
\pgfkeyssetvalue{/sea/factor/theme_nameSnow:Percentage}{8.70802685860151}
\pgfkeyssetvalue{/sea/factor/theme_nameUnderground:Count}{1846678}
\pgfkeyssetvalue{/sea/factor/theme_nameUnderground:Percentage}{6.93986127252349}
\pgfkeyssetvalue{/sea/factor/theme_nameUnderwater:Count}{540587}
\pgfkeyssetvalue{/sea/factor/theme_nameUnderwater:Percentage}{2.03153922109304}
\pgfkeyssetvalue{/sea/factor/game_versionV1.0.0:Count}{72152}
\pgfkeyssetvalue{/sea/factor/game_versionV1.0.0:Percentage}{0.271148987823061}
\pgfkeyssetvalue{/sea/factor/game_versionV1.0.1:Count}{6308783}
\pgfkeyssetvalue{/sea/factor/game_versionV1.0.1:Percentage}{23.7085614375947}
\pgfkeyssetvalue{/sea/factor/game_versionV1.1.0:Count}{1253861}
\pgfkeyssetvalue{/sea/factor/game_versionV1.1.0:Percentage}{4.71204042882818}
\pgfkeyssetvalue{/sea/factor/game_versionV2.0.0:Count}{3977634}
\pgfkeyssetvalue{/sea/factor/game_versionV2.0.0:Percentage}{14.9480462500082}
\pgfkeyssetvalue{/sea/factor/game_versionV3.0.0:Count}{3825551}
\pgfkeyssetvalue{/sea/factor/game_versionV3.0.0:Percentage}{14.3765146013347}
\pgfkeyssetvalue{/sea/factor/game_versionV3.0.1:Count}{11171744}
\pgfkeyssetvalue{/sea/factor/game_versionV3.0.1:Percentage}{41.9836882944112}
\pgfkeyssetvalue{/sea/coefs.both.user/(Intercept)/fml.all.both}{-0.290295313656599}
\pgfkeyssetvalue{/sea/coefs.both.user/difficulty_nameNormal/fml.all.both}{-0.540833733127913}
\pgfkeyssetvalue{/sea/coefs.both.user/difficulty_nameExpert/fml.all.both}{-0.841365306088128}
\pgfkeyssetvalue{/sea/coefs.both.user/difficulty_nameSuper expert/fml.all.both}{-0.976606651955006}
\pgfkeyssetvalue{/sea/coefs.both.user/upload_attempts/fml.all.both}{-0.018743900242282}
\pgfkeyssetvalue{/sea/coefs.both.user/clears/fml.all.both}{0.00471895761659158}
\pgfkeyssetvalue{/sea/coefs.both.user/attempts/fml.all.both}{-0.0044446881921083}
\pgfkeyssetvalue{/sea/coefs.both.user/likes/fml.all.both}{-0.00124465871355162}
\pgfkeyssetvalue{/sea/coefs.both.user/boos/fml.all.both}{-0.00206799699825877}
\pgfkeyssetvalue{/sea/coefs.both.user/unique_players_and_versus/fml.all.both}{0.000479176646489332}
\pgfkeyssetvalue{/sea/coefs.both.user/world_record/fml.all.both}{-0.0029628931257164}
\pgfkeyssetvalue{/sea/coefs.both.user/upload_time/fml.all.both}{-0.000407776441156638}
\pgfkeyssetvalue{/sea/coefs.both.user/num_comments/fml.all.both}{-0.0030938992788645}
\pgfkeyssetvalue{/sea/coefs.both.user/timer/fml.all.both}{0.00021496282020439}
\pgfkeyssetvalue{/sea/coefs.both.user/autoscroll_speed/fml.all.both}{0.0234437863162527}
\pgfkeyssetvalue{/sea/coefs.both.user/gamestyle_nameSMB3/fml.all.both}{-0.0157857657648561}
\pgfkeyssetvalue{/sea/coefs.both.user/gamestyle_nameSMW/fml.all.both}{-0.0236859449943184}
\pgfkeyssetvalue{/sea/coefs.both.user/gamestyle_nameNSMBU/fml.all.both}{-0.0246897205861502}
\pgfkeyssetvalue{/sea/coefs.both.user/gamestyle_nameSM3DW/fml.all.both}{0.0256482326404157}
\pgfkeyssetvalue{/sea/coefs.both.user/theme_nameCastle/fml.all.both}{-0.00710434096975077}
\pgfkeyssetvalue{/sea/coefs.both.user/theme_nameDesert/fml.all.both}{0.0152387128371081}
\pgfkeyssetvalue{/sea/coefs.both.user/theme_nameForest/fml.all.both}{0.0324368612828569}
\pgfkeyssetvalue{/sea/coefs.both.user/theme_nameGhost house/fml.all.both}{0.0211766055310261}
\pgfkeyssetvalue{/sea/coefs.both.user/theme_nameOverworld/fml.all.both}{0.0436967927125345}
\pgfkeyssetvalue{/sea/coefs.both.user/theme_nameSky/fml.all.both}{0.0307906941049028}
\pgfkeyssetvalue{/sea/coefs.both.user/theme_nameSnow/fml.all.both}{0.0332330762549315}
\pgfkeyssetvalue{/sea/coefs.both.user/theme_nameUnderground/fml.all.both}{0.0317361758791437}
\pgfkeyssetvalue{/sea/coefs.both.user/theme_nameUnderwater/fml.all.both}{0.0979670652839371}
\pgfkeyssetvalue{/sea/coefs.both.user/game_versionV1.0.1/fml.all.both}{0.0521051619043287}
\pgfkeyssetvalue{/sea/coefs.both.user/game_versionV1.1.0/fml.all.both}{0.0224892881011931}
\pgfkeyssetvalue{/sea/coefs.both.user/game_versionV2.0.0/fml.all.both}{0.00497806872480644}
\pgfkeyssetvalue{/sea/coefs.both.user/game_versionV3.0.0/fml.all.both}{-0.0149880610381877}
\pgfkeyssetvalue{/sea/coefs.both.user/game_versionV3.0.1/fml.all.both}{-0.0112616499152139}
\pgfkeyssetvalue{/sea/coefs.both.user/(Intercept)/fml.user.all}{-0.289423253172967}
\pgfkeyssetvalue{/sea/coefs.both.user/difficulty_nameNormal/fml.user.all}{-0.53899225017965}
\pgfkeyssetvalue{/sea/coefs.both.user/difficulty_nameExpert/fml.user.all}{-0.840483240044587}
\pgfkeyssetvalue{/sea/coefs.both.user/difficulty_nameSuper expert/fml.user.all}{-0.977017269422785}
\pgfkeyssetvalue{/sea/coefs.both.user/upload_attempts/fml.user.all}{-0.0172500086751786}
\pgfkeyssetvalue{/sea/coefs.both.user/clears/fml.user.all}{0.00470512061074135}
\pgfkeyssetvalue{/sea/coefs.both.user/attempts/fml.user.all}{-0.0043972198349953}
\pgfkeyssetvalue{/sea/coefs.both.user/likes/fml.user.all}{-0.000223454762014352}
\pgfkeyssetvalue{/sea/coefs.both.user/boos/fml.user.all}{0.00141584778367454}
\pgfkeyssetvalue{/sea/coefs.both.user/world_record/fml.user.all}{-0.00320447901056031}
\pgfkeyssetvalue{/sea/coefs.both.user/num_comments/fml.user.all}{-0.00162398877484959}
\pgfkeyssetvalue{/sea/coefs.both.user/autoscroll_speed/fml.user.all}{0.0145812161707053}
\pgfkeyssetvalue{/sea/coefs.both.user/gamestyle_nameSMB3/fml.user.all}{-0.0164493482927988}
\pgfkeyssetvalue{/sea/coefs.both.user/gamestyle_nameSMW/fml.user.all}{-0.0207627515584087}
\pgfkeyssetvalue{/sea/coefs.both.user/gamestyle_nameNSMBU/fml.user.all}{-0.0233503077582833}
\pgfkeyssetvalue{/sea/coefs.both.user/gamestyle_nameSM3DW/fml.user.all}{0.0281192471951481}
\pgfkeyssetvalue{/sea/coefs.both.user/theme_nameCastle/fml.user.all}{-0.00923907367944665}
\pgfkeyssetvalue{/sea/coefs.both.user/theme_nameDesert/fml.user.all}{0.0139118290403921}
\pgfkeyssetvalue{/sea/coefs.both.user/theme_nameForest/fml.user.all}{0.0298820595645692}
\pgfkeyssetvalue{/sea/coefs.both.user/theme_nameGhost house/fml.user.all}{0.0168181788745985}
\pgfkeyssetvalue{/sea/coefs.both.user/theme_nameOverworld/fml.user.all}{0.0388812284335691}
\pgfkeyssetvalue{/sea/coefs.both.user/theme_nameSky/fml.user.all}{0.0233705137738354}
\pgfkeyssetvalue{/sea/coefs.both.user/theme_nameSnow/fml.user.all}{0.0295057517143213}
\pgfkeyssetvalue{/sea/coefs.both.user/theme_nameUnderground/fml.user.all}{0.0276349068672619}
\pgfkeyssetvalue{/sea/coefs.both.user/theme_nameUnderwater/fml.user.all}{0.0986482328166121}
\pgfkeyssetvalue{/sea/coefs.both.user/game_versionV1.0.1/fml.user.all}{0.0501477849181335}
\pgfkeyssetvalue{/sea/coefs.both.user/game_versionV1.1.0/fml.user.all}{0.0082294545336119}
\pgfkeyssetvalue{/sea/coefs.both.user/game_versionV2.0.0/fml.user.all}{-0.00452822958146437}
\pgfkeyssetvalue{/sea/coefs.both.user/game_versionV3.0.0/fml.user.all}{-0.0184803780491776}
\pgfkeyssetvalue{/sea/coefs.both.user/game_versionV3.0.1/fml.user.all}{-0.0154722811569377}
\pgfkeyssetvalue{/sea/coefs.both.user/courses_played/fml.user.all}{6.27850852885636e-06}
\pgfkeyssetvalue{/sea/coefs.both.user/courses_cleared/fml.user.all}{-5.32882748893737e-06}
\pgfkeyssetvalue{/sea/coefs.both.user/courses_attempted/fml.user.all}{2.45344370508604e-06}
\pgfkeyssetvalue{/sea/coefs.both.user/courses_deaths/fml.user.all}{-2.38719467737791e-06}
\pgfkeyssetvalue{/sea/coefs.both.user/maker_points/fml.user.all}{4.68649484717254e-05}
\pgfkeyssetvalue{/sea/coefs.both.user/easy_highscore/fml.user.all}{-3.00485995918809e-06}
\pgfkeyssetvalue{/sea/coefs.both.user/normal_highscore/fml.user.all}{-1.81490359119074e-06}
\pgfkeyssetvalue{/sea/coefs.both.user/expert_highscore/fml.user.all}{-1.68423898204706e-05}
\pgfkeyssetvalue{/sea/coefs.both.user/super_expert_highscore/fml.user.all}{7.45848771990243e-05}
\pgfkeyssetvalue{/sea/coefs.both.user/versus_rating/fml.user.all}{-3.02737005131171e-06}
\pgfkeyssetvalue{/sea/coefs.both.user/first_clears/fml.user.all}{-2.18454907330301e-06}
\pgfkeyssetvalue{/sea/coefs.both.user/world_records/fml.user.all}{-1.78756636828314e-05}
\pgfkeyssetvalue{/sea/coefs.both.user/uploaded_levels/fml.user.all}{-0.000533268348545168}
\pgfkeyssetvalue{/sea/coefs.both.user/comments_enabledTRUE/fml.user.all}{0.00177886912002223}
\pgfkeyssetvalue{/sea/coefs.both.user/regionAmericas/fml.user.all}{-0.00943453608831357}
\pgfkeyssetvalue{/sea/coefs.both.user/regionEurope/fml.user.all}{-0.0167926360173115}
\pgfkeyssetvalue{/sea/coefs.both.user/regionOther/fml.user.all}{-0.00382349490939504}
\pgfkeyssetvalue{/sea/coefs.both.user/likes_maker/fml.user.all}{-2.98733276593577e-06}
\pgfkeyssetvalue{/sea/coefs.df/difficulty_nameNormal/fml.all.attempts}{-0.57192416021069}
\pgfkeyssetvalue{/sea/coefs.df/difficulty_nameExpert/fml.all.attempts}{-0.864410919617049}
\pgfkeyssetvalue{/sea/coefs.df/difficulty_nameSuper expert/fml.all.attempts}{-0.976809259160799}
\pgfkeyssetvalue{/sea/coefs.df/upload_attempts/fml.all.attempts}{-0.0223605449202414}
\pgfkeyssetvalue{/sea/coefs.df/attempts/fml.all.attempts}{-0.00111273531513889}
\pgfkeyssetvalue{/sea/coefs.df/likes/fml.all.attempts}{-0.00172475276192985}
\pgfkeyssetvalue{/sea/coefs.df/boos/fml.all.attempts}{-0.0135800987651035}
\pgfkeyssetvalue{/sea/coefs.df/unique_players_and_versus/fml.all.attempts}{0.0014620582457392}
\pgfkeyssetvalue{/sea/coefs.df/world_record/fml.all.attempts}{-0.00286554759797131}
\pgfkeyssetvalue{/sea/coefs.df/upload_time/fml.all.attempts}{-0.00078134339789504}
\pgfkeyssetvalue{/sea/coefs.df/num_comments/fml.all.attempts}{0.000963023453203649}
\pgfkeyssetvalue{/sea/coefs.df/timer/fml.all.attempts}{0.000248645157757466}
\pgfkeyssetvalue{/sea/coefs.df/autoscroll_speed/fml.all.attempts}{0.0237654178156304}
\pgfkeyssetvalue{/sea/coefs.df/gamestyle_nameSMB3/fml.all.attempts}{-0.0189328524448932}
\pgfkeyssetvalue{/sea/coefs.df/gamestyle_nameSMW/fml.all.attempts}{-0.0275090737918308}
\pgfkeyssetvalue{/sea/coefs.df/gamestyle_nameNSMBU/fml.all.attempts}{-0.0236899266052563}
\pgfkeyssetvalue{/sea/coefs.df/gamestyle_nameSM3DW/fml.all.attempts}{0.0355721952355501}
\pgfkeyssetvalue{/sea/coefs.df/theme_nameCastle/fml.all.attempts}{-0.00855065745501915}
\pgfkeyssetvalue{/sea/coefs.df/theme_nameDesert/fml.all.attempts}{0.0114106699569416}
\pgfkeyssetvalue{/sea/coefs.df/theme_nameForest/fml.all.attempts}{0.0340031064704864}
\pgfkeyssetvalue{/sea/coefs.df/theme_nameGhost house/fml.all.attempts}{0.0310998634608146}
\pgfkeyssetvalue{/sea/coefs.df/theme_nameOverworld/fml.all.attempts}{0.0438851314070066}
\pgfkeyssetvalue{/sea/coefs.df/theme_nameSky/fml.all.attempts}{0.0291483198057378}
\pgfkeyssetvalue{/sea/coefs.df/theme_nameSnow/fml.all.attempts}{0.0330936116388918}
\pgfkeyssetvalue{/sea/coefs.df/theme_nameUnderground/fml.all.attempts}{0.0362669367687585}
\pgfkeyssetvalue{/sea/coefs.df/theme_nameUnderwater/fml.all.attempts}{0.122417813758938}
\pgfkeyssetvalue{/sea/coefs.df/game_versionV1.0.1/fml.all.attempts}{0.0248643446505741}
\pgfkeyssetvalue{/sea/coefs.df/game_versionV1.1.0/fml.all.attempts}{0.0156663528706402}
\pgfkeyssetvalue{/sea/coefs.df/game_versionV2.0.0/fml.all.attempts}{0.00648286883000582}
\pgfkeyssetvalue{/sea/coefs.df/game_versionV3.0.0/fml.all.attempts}{-0.00640885394855772}
\pgfkeyssetvalue{/sea/coefs.df/game_versionV3.0.1/fml.all.attempts}{0.0103961778390094}
\pgfkeyssetvalue{/sea/coefs.df/difficulty_nameNormal/fml.all.clears}{-0.585440475495142}
\pgfkeyssetvalue{/sea/coefs.df/difficulty_nameExpert/fml.all.clears}{-0.877229204887441}
\pgfkeyssetvalue{/sea/coefs.df/difficulty_nameSuper expert/fml.all.clears}{-0.978478308068913}
\pgfkeyssetvalue{/sea/coefs.df/upload_attempts/fml.all.clears}{-0.0226884445323375}
\pgfkeyssetvalue{/sea/coefs.df/clears/fml.all.clears}{5.67274284235531e-06}
\pgfkeyssetvalue{/sea/coefs.df/likes/fml.all.clears}{-8.5652106083689e-05}
\pgfkeyssetvalue{/sea/coefs.df/boos/fml.all.clears}{-0.000176447664158808}
\pgfkeyssetvalue{/sea/coefs.df/unique_players_and_versus/fml.all.clears}{2.20754363544629e-05}
\pgfkeyssetvalue{/sea/coefs.df/world_record/fml.all.clears}{-0.0025216250066995}
\pgfkeyssetvalue{/sea/coefs.df/upload_time/fml.all.clears}{-0.000876226581352668}
\pgfkeyssetvalue{/sea/coefs.df/num_comments/fml.all.clears}{3.21536263037103e-05}
\pgfkeyssetvalue{/sea/coefs.df/timer/fml.all.clears}{0.000267297035997371}
\pgfkeyssetvalue{/sea/coefs.df/autoscroll_speed/fml.all.clears}{0.0261638095219949}
\pgfkeyssetvalue{/sea/coefs.df/gamestyle_nameSMB3/fml.all.clears}{-0.021825700125774}
\pgfkeyssetvalue{/sea/coefs.df/gamestyle_nameSMW/fml.all.clears}{-0.0293880550285828}
\pgfkeyssetvalue{/sea/coefs.df/gamestyle_nameNSMBU/fml.all.clears}{-0.0237992509069036}
\pgfkeyssetvalue{/sea/coefs.df/gamestyle_nameSM3DW/fml.all.clears}{0.0361061277830348}
\pgfkeyssetvalue{/sea/coefs.df/theme_nameCastle/fml.all.clears}{-0.0100234642962215}
\pgfkeyssetvalue{/sea/coefs.df/theme_nameDesert/fml.all.clears}{0.0093393414120122}
\pgfkeyssetvalue{/sea/coefs.df/theme_nameForest/fml.all.clears}{0.0345550476839549}
\pgfkeyssetvalue{/sea/coefs.df/theme_nameGhost house/fml.all.clears}{0.0344245430282033}
\pgfkeyssetvalue{/sea/coefs.df/theme_nameOverworld/fml.all.clears}{0.0387476119497487}
\pgfkeyssetvalue{/sea/coefs.df/theme_nameSky/fml.all.clears}{0.0257188871834229}
\pgfkeyssetvalue{/sea/coefs.df/theme_nameSnow/fml.all.clears}{0.0305010218132584}
\pgfkeyssetvalue{/sea/coefs.df/theme_nameUnderground/fml.all.clears}{0.0369803269779985}
\pgfkeyssetvalue{/sea/coefs.df/theme_nameUnderwater/fml.all.clears}{0.140242104102455}
\pgfkeyssetvalue{/sea/coefs.df/game_versionV1.0.1/fml.all.clears}{-0.00483571546318617}
\pgfkeyssetvalue{/sea/coefs.df/game_versionV1.1.0/fml.all.clears}{0.0061670716074218}
\pgfkeyssetvalue{/sea/coefs.df/game_versionV2.0.0/fml.all.clears}{0.00652215421121083}
\pgfkeyssetvalue{/sea/coefs.df/game_versionV3.0.0/fml.all.clears}{0.00862101856401498}
\pgfkeyssetvalue{/sea/coefs.df/game_versionV3.0.1/fml.all.clears}{0.0333627129737799}
\pgfkeyssetvalue{/sea/coefs.df/difficulty_nameNormal/fml.all}{-0.585472417768336}
\pgfkeyssetvalue{/sea/coefs.df/difficulty_nameExpert/fml.all}{-0.877247615459857}
\pgfkeyssetvalue{/sea/coefs.df/difficulty_nameSuper expert/fml.all}{-0.978485406732147}
\pgfkeyssetvalue{/sea/coefs.df/upload_attempts/fml.all}{-0.0226884221642432}
\pgfkeyssetvalue{/sea/coefs.df/likes/fml.all}{-7.48710430488186e-05}
\pgfkeyssetvalue{/sea/coefs.df/boos/fml.all}{-0.000137135627584439}
\pgfkeyssetvalue{/sea/coefs.df/unique_players_and_versus/fml.all}{2.41009237818002e-05}
\pgfkeyssetvalue{/sea/coefs.df/world_record/fml.all}{-0.00252362333504885}
\pgfkeyssetvalue{/sea/coefs.df/upload_time/fml.all}{-0.000876735141684626}
\pgfkeyssetvalue{/sea/coefs.df/num_comments/fml.all}{4.52777860853182e-05}
\pgfkeyssetvalue{/sea/coefs.df/timer/fml.all}{0.000267179369651149}
\pgfkeyssetvalue{/sea/coefs.df/autoscroll_speed/fml.all}{0.0260603780357469}
\pgfkeyssetvalue{/sea/coefs.df/gamestyle_nameSMB3/fml.all}{-0.021790509793402}
\pgfkeyssetvalue{/sea/coefs.df/gamestyle_nameSMW/fml.all}{-0.029349659520434}
\pgfkeyssetvalue{/sea/coefs.df/gamestyle_nameNSMBU/fml.all}{-0.0237670552310847}
\pgfkeyssetvalue{/sea/coefs.df/gamestyle_nameSM3DW/fml.all}{0.0361384538991556}
\pgfkeyssetvalue{/sea/coefs.df/theme_nameCastle/fml.all}{-0.0100195101488919}
\pgfkeyssetvalue{/sea/coefs.df/theme_nameDesert/fml.all}{0.00934099955315393}
\pgfkeyssetvalue{/sea/coefs.df/theme_nameForest/fml.all}{0.0345693861332028}
\pgfkeyssetvalue{/sea/coefs.df/theme_nameGhost house/fml.all}{0.0344123034710366}
\pgfkeyssetvalue{/sea/coefs.df/theme_nameOverworld/fml.all}{0.038745480094919}
\pgfkeyssetvalue{/sea/coefs.df/theme_nameSky/fml.all}{0.0257132823985571}
\pgfkeyssetvalue{/sea/coefs.df/theme_nameSnow/fml.all}{0.0304920750450994}
\pgfkeyssetvalue{/sea/coefs.df/theme_nameUnderground/fml.all}{0.0369914788041739}
\pgfkeyssetvalue{/sea/coefs.df/theme_nameUnderwater/fml.all}{0.1402577474933}
\pgfkeyssetvalue{/sea/coefs.df/game_versionV1.0.1/fml.all}{-0.00485371622130393}
\pgfkeyssetvalue{/sea/coefs.df/game_versionV1.1.0/fml.all}{0.00618158545461234}
\pgfkeyssetvalue{/sea/coefs.df/game_versionV2.0.0/fml.all}{0.00653007516339166}
\pgfkeyssetvalue{/sea/coefs.df/game_versionV3.0.0/fml.all}{0.00863794346959956}
\pgfkeyssetvalue{/sea/coefs.df/game_versionV3.0.1/fml.all}{0.0333666301452329}
\pgfkeyssetvalue{/sea/coefs.df/difficulty_nameNormal/fml.all.both}{-0.540833733127913}
\pgfkeyssetvalue{/sea/coefs.df/difficulty_nameExpert/fml.all.both}{-0.841365306088128}
\pgfkeyssetvalue{/sea/coefs.df/difficulty_nameSuper expert/fml.all.both}{-0.976606651955006}
\pgfkeyssetvalue{/sea/coefs.df/upload_attempts/fml.all.both}{-0.018743900242282}
\pgfkeyssetvalue{/sea/coefs.df/clears/fml.all.both}{0.00471895761659158}
\pgfkeyssetvalue{/sea/coefs.df/attempts/fml.all.both}{-0.0044446881921083}
\pgfkeyssetvalue{/sea/coefs.df/likes/fml.all.both}{-0.00124465871355162}
\pgfkeyssetvalue{/sea/coefs.df/boos/fml.all.both}{-0.00206799699825877}
\pgfkeyssetvalue{/sea/coefs.df/unique_players_and_versus/fml.all.both}{0.000479176646489332}
\pgfkeyssetvalue{/sea/coefs.df/world_record/fml.all.both}{-0.0029628931257164}
\pgfkeyssetvalue{/sea/coefs.df/upload_time/fml.all.both}{-0.000407776441156638}
\pgfkeyssetvalue{/sea/coefs.df/num_comments/fml.all.both}{-0.0030938992788645}
\pgfkeyssetvalue{/sea/coefs.df/timer/fml.all.both}{0.00021496282020439}
\pgfkeyssetvalue{/sea/coefs.df/autoscroll_speed/fml.all.both}{0.0234437863162527}
\pgfkeyssetvalue{/sea/coefs.df/gamestyle_nameSMB3/fml.all.both}{-0.0157857657648561}
\pgfkeyssetvalue{/sea/coefs.df/gamestyle_nameSMW/fml.all.both}{-0.0236859449943184}
\pgfkeyssetvalue{/sea/coefs.df/gamestyle_nameNSMBU/fml.all.both}{-0.0246897205861502}
\pgfkeyssetvalue{/sea/coefs.df/gamestyle_nameSM3DW/fml.all.both}{0.0256482326404157}
\pgfkeyssetvalue{/sea/coefs.df/theme_nameCastle/fml.all.both}{-0.00710434096975077}
\pgfkeyssetvalue{/sea/coefs.df/theme_nameDesert/fml.all.both}{0.0152387128371081}
\pgfkeyssetvalue{/sea/coefs.df/theme_nameForest/fml.all.both}{0.0324368612828569}
\pgfkeyssetvalue{/sea/coefs.df/theme_nameGhost house/fml.all.both}{0.0211766055310261}
\pgfkeyssetvalue{/sea/coefs.df/theme_nameOverworld/fml.all.both}{0.0436967927125345}
\pgfkeyssetvalue{/sea/coefs.df/theme_nameSky/fml.all.both}{0.0307906941049028}
\pgfkeyssetvalue{/sea/coefs.df/theme_nameSnow/fml.all.both}{0.0332330762549315}
\pgfkeyssetvalue{/sea/coefs.df/theme_nameUnderground/fml.all.both}{0.0317361758791437}
\pgfkeyssetvalue{/sea/coefs.df/theme_nameUnderwater/fml.all.both}{0.0979670652839371}
\pgfkeyssetvalue{/sea/coefs.df/game_versionV1.0.1/fml.all.both}{0.0521051619043287}
\pgfkeyssetvalue{/sea/coefs.df/game_versionV1.1.0/fml.all.both}{0.0224892881011931}
\pgfkeyssetvalue{/sea/coefs.df/game_versionV2.0.0/fml.all.both}{0.00497806872480644}
\pgfkeyssetvalue{/sea/coefs.df/game_versionV3.0.0/fml.all.both}{-0.0149880610381877}
\pgfkeyssetvalue{/sea/coefs.df/game_versionV3.0.1/fml.all.both}{-0.0112616499152139}
\pgfkeyssetvalue{/sea/average.effects/upload_attempts/fml.all.both:mean_delta}{-0.10222727652608}
\pgfkeyssetvalue{/sea/average.effects/upload_attempts/fml.all.both:mean_diff}{5.69916028800128}
\pgfkeyssetvalue{/sea/average.effects/clears/fml.all.both:mean_delta}{0.224960339146886}
\pgfkeyssetvalue{/sea/average.effects/clears/fml.all.both:mean_diff}{43.0999526516648}
\pgfkeyssetvalue{/sea/average.effects/attempts/fml.all.both:mean_delta}{-0.671088097118267}
\pgfkeyssetvalue{/sea/average.effects/attempts/fml.all.both:mean_diff}{249.622084663547}
\pgfkeyssetvalue{/sea/average.effects/likes/fml.all.both:mean_delta}{-0.0108245629006227}
\pgfkeyssetvalue{/sea/average.effects/likes/fml.all.both:mean_diff}{8.73878118316051}
\pgfkeyssetvalue{/sea/average.effects/boos/fml.all.both:mean_delta}{-0.00582270546019859}
\pgfkeyssetvalue{/sea/average.effects/boos/fml.all.both:mean_diff}{2.82093405296234}
\pgfkeyssetvalue{/sea/average.effects/unique_players_and_versus/fml.all.both:mean_delta}{0.0288339887380029}
\pgfkeyssetvalue{/sea/average.effects/unique_players_and_versus/fml.all.both:mean_diff}{59.3370345761482}
\pgfkeyssetvalue{/sea/average.effects/world_record/fml.all.both:mean_delta}{-0.120705275937001}
\pgfkeyssetvalue{/sea/average.effects/world_record/fml.all.both:mean_diff}{43.351034609338}
\pgfkeyssetvalue{/sea/average.effects/upload_time/fml.all.both:mean_delta}{-0.0209181925777576}
\pgfkeyssetvalue{/sea/average.effects/upload_time/fml.all.both:mean_diff}{51.8317516037236}
\pgfkeyssetvalue{/sea/average.effects/num_comments/fml.all.both:mean_delta}{-0.00586287082201631}
\pgfkeyssetvalue{/sea/average.effects/num_comments/fml.all.both:mean_diff}{1.8976131532365}
\pgfkeyssetvalue{/sea/average.effects/timer/fml.all.both:mean_delta}{0.0211647912653332}
\pgfkeyssetvalue{/sea/average.effects/timer/fml.all.both:mean_diff}{97.440938707344}
\pgfkeyssetvalue{/sea/average.effects/autoscroll_speed/fml.all.both:mean_delta}{0.0006277767462044}
\pgfkeyssetvalue{/sea/average.effects/autoscroll_speed/fml.all.both:mean_diff}{0.0270821350960989}
\pgfkeyssetvalue{/sea/average.effects/upload_attempts/fml.user.all:mean_delta}{-0.0944097813616092}
\pgfkeyssetvalue{/sea/average.effects/upload_attempts/fml.user.all:mean_diff}{5.69916028800128}
\pgfkeyssetvalue{/sea/average.effects/clears/fml.user.all:mean_delta}{0.224233446205329}
\pgfkeyssetvalue{/sea/average.effects/clears/fml.user.all:mean_diff}{43.0999526516648}
\pgfkeyssetvalue{/sea/average.effects/attempts/fml.user.all:mean_delta}{-0.667150075925706}
\pgfkeyssetvalue{/sea/average.effects/attempts/fml.user.all:mean_diff}{249.622084663547}
\pgfkeyssetvalue{/sea/average.effects/likes/fml.user.all:mean_delta}{-0.00195103472718716}
\pgfkeyssetvalue{/sea/average.effects/likes/fml.user.all:mean_diff}{8.73878118316051}
\pgfkeyssetvalue{/sea/average.effects/boos/fml.user.all:mean_delta}{0.00399916383486754}
\pgfkeyssetvalue{/sea/average.effects/boos/fml.user.all:mean_diff}{2.82093405296234}
\pgfkeyssetvalue{/sea/average.effects/world_record/fml.user.all:mean_delta}{-0.12989426149971}
\pgfkeyssetvalue{/sea/average.effects/world_record/fml.user.all:mean_diff}{43.351034609338}
\pgfkeyssetvalue{/sea/average.effects/num_comments/fml.user.all:mean_delta}{-0.00307945621539152}
\pgfkeyssetvalue{/sea/average.effects/num_comments/fml.user.all:mean_diff}{1.8976131532365}
\pgfkeyssetvalue{/sea/average.effects/autoscroll_speed/fml.user.all:mean_delta}{0.000392116015575095}
\pgfkeyssetvalue{/sea/average.effects/autoscroll_speed/fml.user.all:mean_diff}{0.0270821350960989}
\pgfkeyssetvalue{/sea/average.effects/courses_played/fml.user.all:mean_delta}{0.00632255085270628}
\pgfkeyssetvalue{/sea/average.effects/courses_played/fml.user.all:mean_diff}{1003.84783017433}
\pgfkeyssetvalue{/sea/average.effects/courses_cleared/fml.user.all:mean_delta}{-0.00351308085472368}
\pgfkeyssetvalue{/sea/average.effects/courses_cleared/fml.user.all:mean_diff}{660.418608874039}
\pgfkeyssetvalue{/sea/average.effects/courses_attempted/fml.user.all:mean_delta}{0.0118991828156407}
\pgfkeyssetvalue{/sea/average.effects/courses_attempted/fml.user.all:mean_diff}{4821.36951637481}
\pgfkeyssetvalue{/sea/average.effects/courses_deaths/fml.user.all:mean_delta}{-0.00970539899020273}
\pgfkeyssetvalue{/sea/average.effects/courses_deaths/fml.user.all:mean_diff}{4085.46139726807}
\pgfkeyssetvalue{/sea/average.effects/maker_points/fml.user.all:mean_delta}{0.0722437006183874}
\pgfkeyssetvalue{/sea/average.effects/maker_points/fml.user.all:mean_diff}{1488.42591819369}
\pgfkeyssetvalue{/sea/average.effects/easy_highscore/fml.user.all:mean_delta}{-0.000586149181218021}
\pgfkeyssetvalue{/sea/average.effects/easy_highscore/fml.user.all:mean_diff}{195.123952817593}
\pgfkeyssetvalue{/sea/average.effects/normal_highscore/fml.user.all:mean_delta}{-9.01575820871914e-05}
\pgfkeyssetvalue{/sea/average.effects/normal_highscore/fml.user.all:mean_diff}{49.6784320371207}
\pgfkeyssetvalue{/sea/average.effects/expert_highscore/fml.user.all:mean_delta}{-0.000154757307883724}
\pgfkeyssetvalue{/sea/average.effects/expert_highscore/fml.user.all:mean_diff}{9.1891935963507}
\pgfkeyssetvalue{/sea/average.effects/super_expert_highscore/fml.user.all:mean_delta}{4.79017446519681e-05}
\pgfkeyssetvalue{/sea/average.effects/super_expert_highscore/fml.user.all:mean_diff}{0.64225330263758}
\pgfkeyssetvalue{/sea/average.effects/versus_rating/fml.user.all:mean_delta}{-0.00396168461008828}
\pgfkeyssetvalue{/sea/average.effects/versus_rating/fml.user.all:mean_diff}{1311.21957162346}
\pgfkeyssetvalue{/sea/average.effects/first_clears/fml.user.all:mean_delta}{-6.89922776682117e-05}
\pgfkeyssetvalue{/sea/average.effects/first_clears/fml.user.all:mean_diff}{31.5829858101682}
\pgfkeyssetvalue{/sea/average.effects/world_records/fml.user.all:mean_delta}{-0.000588279049220763}
\pgfkeyssetvalue{/sea/average.effects/world_records/fml.user.all:mean_diff}{32.9188836941522}
\pgfkeyssetvalue{/sea/average.effects/uploaded_levels/fml.user.all:mean_delta}{-0.0169505715611721}
\pgfkeyssetvalue{/sea/average.effects/uploaded_levels/fml.user.all:mean_diff}{32.050126622664}
\pgfkeyssetvalue{/sea/average.effects/likes_maker/fml.user.all:mean_delta}{-0.00105721912502821}
\pgfkeyssetvalue{/sea/average.effects/likes_maker/fml.user.all:mean_diff}{354.087367655122}
\pgfkeyssetvalue{/sea/N.99.players}{267089}
\pgfkeyssetvalue{/sea/N.99.9.players}{26610}
\pgfkeyssetvalue{/sea/N.99.makers}{266120}
\pgfkeyssetvalue{/sea/N.99.9.makers}{26678}
\pgfkeyssetvalue{/sea/N.99.9.both}{12748}
\pgfkeyssetvalue{/sea/N.99.9.both.percent.players}{0.479068019541526}
\pgfkeyssetvalue{/sea/N.99.9.both.percent.makers}{0.477846915061099}
\pgfkeyssetvalue{/sea/diff.coefs/difficulty_nameNormal:fml.user.all:data_99_9_players}{0.0528866789590141}
\pgfkeyssetvalue{/sea/diff.coefs/difficulty_nameExpert:fml.user.all:data_99_9_players}{0.0790332222992949}
\pgfkeyssetvalue{/sea/diff.coefs/difficulty_nameSuper expert:fml.user.all:data_99_9_players}{0.0755996816970362}
\pgfkeyssetvalue{/sea/diff.coefs/upload_attempts:fml.user.all:data_99_9_players}{-0.000205382062725357}
\pgfkeyssetvalue{/sea/diff.coefs/clears:fml.user.all:data_99_9_players}{-0.00467674328409418}
\pgfkeyssetvalue{/sea/diff.coefs/attempts:fml.user.all:data_99_9_players}{0.00437042640730967}
\pgfkeyssetvalue{/sea/diff.coefs/likes:fml.user.all:data_99_9_players}{0.000241191979153843}
\pgfkeyssetvalue{/sea/diff.coefs/boos:fml.user.all:data_99_9_players}{-0.0013780311448095}
\pgfkeyssetvalue{/sea/diff.coefs/world_record:fml.user.all:data_99_9_players}{0.00334693467669123}
\pgfkeyssetvalue{/sea/diff.coefs/num_comments:fml.user.all:data_99_9_players}{0.00159824191071245}
\pgfkeyssetvalue{/sea/diff.coefs/autoscroll_speed:fml.user.all:data_99_9_players}{0.010449754091449}
\pgfkeyssetvalue{/sea/diff.coefs/gamestyle_nameSMB3:fml.user.all:data_99_9_players}{-0.0455412827911497}
\pgfkeyssetvalue{/sea/diff.coefs/gamestyle_nameSMW:fml.user.all:data_99_9_players}{-0.0347439615210403}
\pgfkeyssetvalue{/sea/diff.coefs/gamestyle_nameNSMBU:fml.user.all:data_99_9_players}{-0.0986516138298517}
\pgfkeyssetvalue{/sea/diff.coefs/gamestyle_nameSM3DW:fml.user.all:data_99_9_players}{-0.0660449489246109}
\pgfkeyssetvalue{/sea/diff.coefs/theme_nameCastle:fml.user.all:data_99_9_players}{0.019895235498894}
\pgfkeyssetvalue{/sea/diff.coefs/theme_nameDesert:fml.user.all:data_99_9_players}{-0.0631848279667551}
\pgfkeyssetvalue{/sea/diff.coefs/theme_nameForest:fml.user.all:data_99_9_players}{-0.0606767904590123}
\pgfkeyssetvalue{/sea/diff.coefs/theme_nameGhost house:fml.user.all:data_99_9_players}{0.00889742606732691}
\pgfkeyssetvalue{/sea/diff.coefs/theme_nameOverworld:fml.user.all:data_99_9_players}{-0.0418571053377779}
\pgfkeyssetvalue{/sea/diff.coefs/theme_nameSky:fml.user.all:data_99_9_players}{-0.000440463307426864}
\pgfkeyssetvalue{/sea/diff.coefs/theme_nameSnow:fml.user.all:data_99_9_players}{-0.0487317440741117}
\pgfkeyssetvalue{/sea/diff.coefs/theme_nameUnderground:fml.user.all:data_99_9_players}{0.000850587039265882}
\pgfkeyssetvalue{/sea/diff.coefs/theme_nameUnderwater:fml.user.all:data_99_9_players}{-0.0807247974574896}
\pgfkeyssetvalue{/sea/diff.coefs/game_versionV1.0.1:fml.user.all:data_99_9_players}{0.0393127927942853}
\pgfkeyssetvalue{/sea/diff.coefs/game_versionV1.1.0:fml.user.all:data_99_9_players}{0.0695715812968338}
\pgfkeyssetvalue{/sea/diff.coefs/game_versionV2.0.0:fml.user.all:data_99_9_players}{0.126873680389959}
\pgfkeyssetvalue{/sea/diff.coefs/game_versionV3.0.0:fml.user.all:data_99_9_players}{0.128606059522092}
\pgfkeyssetvalue{/sea/diff.coefs/game_versionV3.0.1:fml.user.all:data_99_9_players}{0.16331136563896}
\pgfkeyssetvalue{/sea/diff.coefs/courses_played:fml.user.all:data_99_9_players}{-5.24624792519468e-06}
\pgfkeyssetvalue{/sea/diff.coefs/courses_cleared:fml.user.all:data_99_9_players}{-2.65695527534859e-06}
\pgfkeyssetvalue{/sea/diff.coefs/courses_attempted:fml.user.all:data_99_9_players}{4.30026595821253e-06}
\pgfkeyssetvalue{/sea/diff.coefs/courses_deaths:fml.user.all:data_99_9_players}{-4.36162531580742e-06}
\pgfkeyssetvalue{/sea/diff.coefs/maker_points:fml.user.all:data_99_9_players}{-4.0880457869763e-05}
\pgfkeyssetvalue{/sea/diff.coefs/easy_highscore:fml.user.all:data_99_9_players}{4.32048725318435e-06}
\pgfkeyssetvalue{/sea/diff.coefs/normal_highscore:fml.user.all:data_99_9_players}{-5.78351677549271e-07}
\pgfkeyssetvalue{/sea/diff.coefs/expert_highscore:fml.user.all:data_99_9_players}{1.47295184740326e-05}
\pgfkeyssetvalue{/sea/diff.coefs/super_expert_highscore:fml.user.all:data_99_9_players}{-7.09224602257752e-05}
\pgfkeyssetvalue{/sea/diff.coefs/versus_rating:fml.user.all:data_99_9_players}{8.56985789976505e-06}
\pgfkeyssetvalue{/sea/diff.coefs/first_clears:fml.user.all:data_99_9_players}{-1.28669009624716e-06}
\pgfkeyssetvalue{/sea/diff.coefs/world_records:fml.user.all:data_99_9_players}{1.19375085009255e-05}
\pgfkeyssetvalue{/sea/diff.coefs/uploaded_levels:fml.user.all:data_99_9_players}{-0.000192331976914928}
\pgfkeyssetvalue{/sea/diff.coefs/comments_enabledTRUE:fml.user.all:data_99_9_players}{-0.00178086680238176}
\pgfkeyssetvalue{/sea/diff.coefs/regionAmericas:fml.user.all:data_99_9_players}{-0.0546440287116363}
\pgfkeyssetvalue{/sea/diff.coefs/regionEurope:fml.user.all:data_99_9_players}{-0.0448717456109428}
\pgfkeyssetvalue{/sea/diff.coefs/regionOther:fml.user.all:data_99_9_players}{-0.0609841855548167}
\pgfkeyssetvalue{/sea/diff.coefs/likes_maker:fml.user.all:data_99_9_players}{3.0722251599391e-06}
\pgfkeyssetvalue{/sea/diff.coefs/difficulty_nameNormal:fml.all.both:data_99_9_players}{0.0494301137602555}
\pgfkeyssetvalue{/sea/diff.coefs/difficulty_nameExpert:fml.all.both:data_99_9_players}{0.0639291584747207}
\pgfkeyssetvalue{/sea/diff.coefs/difficulty_nameSuper expert:fml.all.both:data_99_9_players}{0.0399496614922541}
\pgfkeyssetvalue{/sea/diff.coefs/upload_attempts:fml.all.both:data_99_9_players}{-0.0044526692462411}
\pgfkeyssetvalue{/sea/diff.coefs/clears:fml.all.both:data_99_9_players}{-0.00469442573594758}
\pgfkeyssetvalue{/sea/diff.coefs/attempts:fml.all.both:data_99_9_players}{0.00442193709685446}
\pgfkeyssetvalue{/sea/diff.coefs/likes:fml.all.both:data_99_9_players}{0.00123041122198309}
\pgfkeyssetvalue{/sea/diff.coefs/boos:fml.all.both:data_99_9_players}{0.00201105918539413}
\pgfkeyssetvalue{/sea/diff.coefs/unique_players_and_versus:fml.all.both:data_99_9_players}{-0.000468859086886209}
\pgfkeyssetvalue{/sea/diff.coefs/world_record:fml.all.both:data_99_9_players}{0.00470442697478513}
\pgfkeyssetvalue{/sea/diff.coefs/upload_time:fml.all.both:data_99_9_players}{-0.00138303952452368}
\pgfkeyssetvalue{/sea/diff.coefs/num_comments:fml.all.both:data_99_9_players}{0.00307497222819175}
\pgfkeyssetvalue{/sea/diff.coefs/timer:fml.all.both:data_99_9_players}{0.000112201882302321}
\pgfkeyssetvalue{/sea/diff.coefs/autoscroll_speed:fml.all.both:data_99_9_players}{-0.00308377716664188}
\pgfkeyssetvalue{/sea/diff.coefs/gamestyle_nameSMB3:fml.all.both:data_99_9_players}{-0.054867536097538}
\pgfkeyssetvalue{/sea/diff.coefs/gamestyle_nameSMW:fml.all.both:data_99_9_players}{-0.0389383549483248}
\pgfkeyssetvalue{/sea/diff.coefs/gamestyle_nameNSMBU:fml.all.both:data_99_9_players}{-0.103410291480613}
\pgfkeyssetvalue{/sea/diff.coefs/gamestyle_nameSM3DW:fml.all.both:data_99_9_players}{-0.0673877104199621}
\pgfkeyssetvalue{/sea/diff.coefs/theme_nameCastle:fml.all.both:data_99_9_players}{0.022857239825958}
\pgfkeyssetvalue{/sea/diff.coefs/theme_nameDesert:fml.all.both:data_99_9_players}{-0.06822691279655}
\pgfkeyssetvalue{/sea/diff.coefs/theme_nameForest:fml.all.both:data_99_9_players}{-0.0684664078443924}
\pgfkeyssetvalue{/sea/diff.coefs/theme_nameGhost house:fml.all.both:data_99_9_players}{-0.000556723748857069}
\pgfkeyssetvalue{/sea/diff.coefs/theme_nameOverworld:fml.all.both:data_99_9_players}{-0.0591474022752798}
\pgfkeyssetvalue{/sea/diff.coefs/theme_nameSky:fml.all.both:data_99_9_players}{-0.0100525026835021}
\pgfkeyssetvalue{/sea/diff.coefs/theme_nameSnow:fml.all.both:data_99_9_players}{-0.0479439487723222}
\pgfkeyssetvalue{/sea/diff.coefs/theme_nameUnderground:fml.all.both:data_99_9_players}{-0.00117929830916341}
\pgfkeyssetvalue{/sea/diff.coefs/theme_nameUnderwater:fml.all.both:data_99_9_players}{-0.087897650497252}
\pgfkeyssetvalue{/sea/diff.coefs/game_versionV1.0.1:fml.all.both:data_99_9_players}{0.0340655358629429}
\pgfkeyssetvalue{/sea/diff.coefs/game_versionV1.1.0:fml.all.both:data_99_9_players}{0.0570078962438709}
\pgfkeyssetvalue{/sea/diff.coefs/game_versionV2.0.0:fml.all.both:data_99_9_players}{0.110202803283631}
\pgfkeyssetvalue{/sea/diff.coefs/game_versionV3.0.0:fml.all.both:data_99_9_players}{0.110745565642423}
\pgfkeyssetvalue{/sea/diff.coefs/game_versionV3.0.1:fml.all.both:data_99_9_players}{0.171555942033137}
\pgfkeyssetvalue{/sea/diff.coefs/difficulty_nameNormal:fml.user.all:data_99_9_makers}{-0.0409566585900923}
\pgfkeyssetvalue{/sea/diff.coefs/difficulty_nameExpert:fml.user.all:data_99_9_makers}{-0.0176251035347488}
\pgfkeyssetvalue{/sea/diff.coefs/difficulty_nameSuper expert:fml.user.all:data_99_9_makers}{0.00904840872214796}
\pgfkeyssetvalue{/sea/diff.coefs/upload_attempts:fml.user.all:data_99_9_makers}{-0.00643772176535096}
\pgfkeyssetvalue{/sea/diff.coefs/clears:fml.user.all:data_99_9_makers}{-0.00468398358780209}
\pgfkeyssetvalue{/sea/diff.coefs/attempts:fml.user.all:data_99_9_makers}{0.00437609723822274}
\pgfkeyssetvalue{/sea/diff.coefs/likes:fml.user.all:data_99_9_makers}{0.0002404614400775}
\pgfkeyssetvalue{/sea/diff.coefs/boos:fml.user.all:data_99_9_makers}{-0.00131836691873843}
\pgfkeyssetvalue{/sea/diff.coefs/world_record:fml.user.all:data_99_9_makers}{0.00296945406310878}
\pgfkeyssetvalue{/sea/diff.coefs/num_comments:fml.user.all:data_99_9_makers}{0.00160398617822577}
\pgfkeyssetvalue{/sea/diff.coefs/autoscroll_speed:fml.user.all:data_99_9_makers}{0.00919612727101193}
\pgfkeyssetvalue{/sea/diff.coefs/gamestyle_nameSMB3:fml.user.all:data_99_9_makers}{-0.0886741799444913}
\pgfkeyssetvalue{/sea/diff.coefs/gamestyle_nameSMW:fml.user.all:data_99_9_makers}{-0.0894073192792681}
\pgfkeyssetvalue{/sea/diff.coefs/gamestyle_nameNSMBU:fml.user.all:data_99_9_makers}{-0.122033562928283}
\pgfkeyssetvalue{/sea/diff.coefs/gamestyle_nameSM3DW:fml.user.all:data_99_9_makers}{-0.0786751824721216}
\pgfkeyssetvalue{/sea/diff.coefs/theme_nameCastle:fml.user.all:data_99_9_makers}{0.0516119223701673}
\pgfkeyssetvalue{/sea/diff.coefs/theme_nameDesert:fml.user.all:data_99_9_makers}{0.00347548855052393}
\pgfkeyssetvalue{/sea/diff.coefs/theme_nameForest:fml.user.all:data_99_9_makers}{-0.00236022447927553}
\pgfkeyssetvalue{/sea/diff.coefs/theme_nameGhost house:fml.user.all:data_99_9_makers}{0.0459160044360305}
\pgfkeyssetvalue{/sea/diff.coefs/theme_nameOverworld:fml.user.all:data_99_9_makers}{0.0451789367191036}
\pgfkeyssetvalue{/sea/diff.coefs/theme_nameSky:fml.user.all:data_99_9_makers}{0.0429560864141378}
\pgfkeyssetvalue{/sea/diff.coefs/theme_nameSnow:fml.user.all:data_99_9_makers}{0.00422287568158652}
\pgfkeyssetvalue{/sea/diff.coefs/theme_nameUnderground:fml.user.all:data_99_9_makers}{0.0615572498569661}
\pgfkeyssetvalue{/sea/diff.coefs/theme_nameUnderwater:fml.user.all:data_99_9_makers}{-0.0554172488438824}
\pgfkeyssetvalue{/sea/diff.coefs/game_versionV1.0.1:fml.user.all:data_99_9_makers}{0.0812549205707349}
\pgfkeyssetvalue{/sea/diff.coefs/game_versionV1.1.0:fml.user.all:data_99_9_makers}{0.11535690216344}
\pgfkeyssetvalue{/sea/diff.coefs/game_versionV2.0.0:fml.user.all:data_99_9_makers}{0.15560255084332}
\pgfkeyssetvalue{/sea/diff.coefs/game_versionV3.0.0:fml.user.all:data_99_9_makers}{0.164839540688819}
\pgfkeyssetvalue{/sea/diff.coefs/game_versionV3.0.1:fml.user.all:data_99_9_makers}{0.184038271248051}
\pgfkeyssetvalue{/sea/diff.coefs/courses_played:fml.user.all:data_99_9_makers}{-6.43741242456475e-06}
\pgfkeyssetvalue{/sea/diff.coefs/courses_cleared:fml.user.all:data_99_9_makers}{-4.45107095248254e-06}
\pgfkeyssetvalue{/sea/diff.coefs/courses_attempted:fml.user.all:data_99_9_makers}{5.17236120001208e-06}
\pgfkeyssetvalue{/sea/diff.coefs/courses_deaths:fml.user.all:data_99_9_makers}{-5.40943624216528e-06}
\pgfkeyssetvalue{/sea/diff.coefs/maker_points:fml.user.all:data_99_9_makers}{-2.75173517489957e-05}
\pgfkeyssetvalue{/sea/diff.coefs/easy_highscore:fml.user.all:data_99_9_makers}{7.9917895438042e-06}
\pgfkeyssetvalue{/sea/diff.coefs/normal_highscore:fml.user.all:data_99_9_makers}{2.40492973624384e-06}
\pgfkeyssetvalue{/sea/diff.coefs/expert_highscore:fml.user.all:data_99_9_makers}{1.52251972700412e-05}
\pgfkeyssetvalue{/sea/diff.coefs/super_expert_highscore:fml.user.all:data_99_9_makers}{-6.10824674689425e-05}
\pgfkeyssetvalue{/sea/diff.coefs/versus_rating:fml.user.all:data_99_9_makers}{6.09486659675085e-06}
\pgfkeyssetvalue{/sea/diff.coefs/first_clears:fml.user.all:data_99_9_makers}{-2.1107298376899e-06}
\pgfkeyssetvalue{/sea/diff.coefs/world_records:fml.user.all:data_99_9_makers}{1.41616473856576e-05}
\pgfkeyssetvalue{/sea/diff.coefs/uploaded_levels:fml.user.all:data_99_9_makers}{0.00019250157567885}
\pgfkeyssetvalue{/sea/diff.coefs/comments_enabledTRUE:fml.user.all:data_99_9_makers}{-0.0351276646823313}
\pgfkeyssetvalue{/sea/diff.coefs/regionAmericas:fml.user.all:data_99_9_makers}{-0.0471200611709415}
\pgfkeyssetvalue{/sea/diff.coefs/regionEurope:fml.user.all:data_99_9_makers}{-0.0520860126485244}
\pgfkeyssetvalue{/sea/diff.coefs/regionOther:fml.user.all:data_99_9_makers}{-0.0977833851334055}
\pgfkeyssetvalue{/sea/diff.coefs/likes_maker:fml.user.all:data_99_9_makers}{2.91997508983233e-06}
\pgfkeyssetvalue{/sea/diff.coefs/difficulty_nameNormal:fml.all.both:data_99_9_makers}{-0.0404375058845632}
\pgfkeyssetvalue{/sea/diff.coefs/difficulty_nameExpert:fml.all.both:data_99_9_makers}{-0.0188165059583724}
\pgfkeyssetvalue{/sea/diff.coefs/difficulty_nameSuper expert:fml.all.both:data_99_9_makers}{0.0087064440891601}
\pgfkeyssetvalue{/sea/diff.coefs/upload_attempts:fml.all.both:data_99_9_makers}{-0.011444212200566}
\pgfkeyssetvalue{/sea/diff.coefs/clears:fml.all.both:data_99_9_makers}{-0.00470184550982422}
\pgfkeyssetvalue{/sea/diff.coefs/attempts:fml.all.both:data_99_9_makers}{0.00442809743030037}
\pgfkeyssetvalue{/sea/diff.coefs/likes:fml.all.both:data_99_9_makers}{0.00122255579178077}
\pgfkeyssetvalue{/sea/diff.coefs/boos:fml.all.both:data_99_9_makers}{0.0020573334870736}
\pgfkeyssetvalue{/sea/diff.coefs/unique_players_and_versus:fml.all.both:data_99_9_makers}{-0.000467204587460746}
\pgfkeyssetvalue{/sea/diff.coefs/world_record:fml.all.both:data_99_9_makers}{0.00318585737188126}
\pgfkeyssetvalue{/sea/diff.coefs/upload_time:fml.all.both:data_99_9_makers}{-0.000754417102205251}
\pgfkeyssetvalue{/sea/diff.coefs/num_comments:fml.all.both:data_99_9_makers}{0.00308370376718881}
\pgfkeyssetvalue{/sea/diff.coefs/timer:fml.all.both:data_99_9_makers}{0.000301857094757274}
\pgfkeyssetvalue{/sea/diff.coefs/autoscroll_speed:fml.all.both:data_99_9_makers}{-0.0112085103432775}
\pgfkeyssetvalue{/sea/diff.coefs/gamestyle_nameSMB3:fml.all.both:data_99_9_makers}{-0.100170729615353}
\pgfkeyssetvalue{/sea/diff.coefs/gamestyle_nameSMW:fml.all.both:data_99_9_makers}{-0.0858395847210125}
\pgfkeyssetvalue{/sea/diff.coefs/gamestyle_nameNSMBU:fml.all.both:data_99_9_makers}{-0.11760922416708}
\pgfkeyssetvalue{/sea/diff.coefs/gamestyle_nameSM3DW:fml.all.both:data_99_9_makers}{-0.0769065086703223}
\pgfkeyssetvalue{/sea/diff.coefs/theme_nameCastle:fml.all.both:data_99_9_makers}{0.0436459915922939}
\pgfkeyssetvalue{/sea/diff.coefs/theme_nameDesert:fml.all.both:data_99_9_makers}{-0.0112606676841145}
\pgfkeyssetvalue{/sea/diff.coefs/theme_nameForest:fml.all.both:data_99_9_makers}{-0.0223157991744913}
\pgfkeyssetvalue{/sea/diff.coefs/theme_nameGhost house:fml.all.both:data_99_9_makers}{0.0224041719590411}
\pgfkeyssetvalue{/sea/diff.coefs/theme_nameOverworld:fml.all.both:data_99_9_makers}{0.026052252408912}
\pgfkeyssetvalue{/sea/diff.coefs/theme_nameSky:fml.all.both:data_99_9_makers}{0.0281325460933466}
\pgfkeyssetvalue{/sea/diff.coefs/theme_nameSnow:fml.all.both:data_99_9_makers}{0.00682166280507146}
\pgfkeyssetvalue{/sea/diff.coefs/theme_nameUnderground:fml.all.both:data_99_9_makers}{0.0629883042121917}
\pgfkeyssetvalue{/sea/diff.coefs/theme_nameUnderwater:fml.all.both:data_99_9_makers}{-0.060177358676502}
\pgfkeyssetvalue{/sea/diff.coefs/game_versionV1.0.1:fml.all.both:data_99_9_makers}{0.134121030276683}
\pgfkeyssetvalue{/sea/diff.coefs/game_versionV1.1.0:fml.all.both:data_99_9_makers}{0.133413003071847}
\pgfkeyssetvalue{/sea/diff.coefs/game_versionV2.0.0:fml.all.both:data_99_9_makers}{0.173387785122898}
\pgfkeyssetvalue{/sea/diff.coefs/game_versionV3.0.0:fml.all.both:data_99_9_makers}{0.184897629311853}
\pgfkeyssetvalue{/sea/diff.coefs/game_versionV3.0.1:fml.all.both:data_99_9_makers}{0.225088729278277}
\pgfkeyssetvalue{/sea/insignificant.vars/data_99_9_makers:fml.all.both:N_vars}{11}
\pgfkeyssetvalue{/sea/insignificant.vars/data_99_9_makers:fml.user.all:N_vars}{26}
\pgfkeyssetvalue{/sea/insignificant.vars/data_99_9_players:fml.all.both:N_vars}{13}
\pgfkeyssetvalue{/sea/insignificant.vars/data_99_9_players:fml.user.all:N_vars}{25}
\pgfkeyssetvalue{/sea/N.fml.causal.variables}{4}
\pgfkeyssetvalue{/sea/coefs.causal/(Intercept)/fml.causal}{-0.506321105127101}
\pgfkeyssetvalue{/sea/coefs.causal/gamestyle_nameSMB3/fml.causal}{-0.0681488942702818}
\pgfkeyssetvalue{/sea/coefs.causal/gamestyle_nameSMW/fml.causal}{-0.117399930025597}
\pgfkeyssetvalue{/sea/coefs.causal/gamestyle_nameNSMBU/fml.causal}{-0.0509237382822227}
\pgfkeyssetvalue{/sea/coefs.causal/gamestyle_nameSM3DW/fml.causal}{0.126116049677774}
\pgfkeyssetvalue{/sea/coefs.causal/theme_nameCastle/fml.causal}{-0.0523856832397982}
\pgfkeyssetvalue{/sea/coefs.causal/theme_nameDesert/fml.causal}{0.0122647065868058}
\pgfkeyssetvalue{/sea/coefs.causal/theme_nameForest/fml.causal}{0.0827737015917083}
\pgfkeyssetvalue{/sea/coefs.causal/theme_nameGhost house/fml.causal}{-0.0226463920871497}
\pgfkeyssetvalue{/sea/coefs.causal/theme_nameOverworld/fml.causal}{0.142302419242461}
\pgfkeyssetvalue{/sea/coefs.causal/theme_nameSky/fml.causal}{0.0509661615946526}
\pgfkeyssetvalue{/sea/coefs.causal/theme_nameSnow/fml.causal}{0.0694591481568083}
\pgfkeyssetvalue{/sea/coefs.causal/theme_nameUnderground/fml.causal}{0.0636468325460664}
\pgfkeyssetvalue{/sea/coefs.causal/theme_nameUnderwater/fml.causal}{0.378275368011175}
\pgfkeyssetvalue{/sea/coefs.causal/game_versionV1.0.1/fml.causal}{-0.127304156773228}
\pgfkeyssetvalue{/sea/coefs.causal/game_versionV1.1.0/fml.causal}{-0.0390706154625893}
\pgfkeyssetvalue{/sea/coefs.causal/game_versionV2.0.0/fml.causal}{-0.0166927793329884}
\pgfkeyssetvalue{/sea/coefs.causal/game_versionV3.0.0/fml.causal}{0.0121595792293201}
\pgfkeyssetvalue{/sea/coefs.causal/game_versionV3.0.1/fml.causal}{0.0580085719760945}
\pgfkeyssetvalue{/sea/coefs.causal/upload_attempts/fml.causal}{-0.191122821556644}
\pgfkeyssetvalue{/sea/causal.diff/gamestyle_nameSMB3:fml.all.both}{-0.0523631285054257}
\pgfkeyssetvalue{/sea/causal.diff/gamestyle_nameSMB3:fml.user.all}{-0.0516995459774831}
\pgfkeyssetvalue{/sea/causal.diff/gamestyle_nameSMW:fml.all.both}{-0.0937139850312786}
\pgfkeyssetvalue{/sea/causal.diff/gamestyle_nameSMW:fml.user.all}{-0.0966371784671883}
\pgfkeyssetvalue{/sea/causal.diff/gamestyle_nameNSMBU:fml.all.both}{-0.0262340176960726}
\pgfkeyssetvalue{/sea/causal.diff/gamestyle_nameNSMBU:fml.user.all}{-0.0275734305239395}
\pgfkeyssetvalue{/sea/causal.diff/gamestyle_nameSM3DW:fml.all.both}{0.100467817037359}
\pgfkeyssetvalue{/sea/causal.diff/gamestyle_nameSM3DW:fml.user.all}{0.0979968024826263}
\pgfkeyssetvalue{/sea/causal.diff/theme_nameCastle:fml.all.both}{-0.0452813422700474}
\pgfkeyssetvalue{/sea/causal.diff/theme_nameCastle:fml.user.all}{-0.0431466095603515}
\pgfkeyssetvalue{/sea/causal.diff/theme_nameDesert:fml.all.both}{-0.00297400625030231}
\pgfkeyssetvalue{/sea/causal.diff/theme_nameDesert:fml.user.all}{-0.00164712245358634}
\pgfkeyssetvalue{/sea/causal.diff/theme_nameForest:fml.all.both}{0.0503368403088513}
\pgfkeyssetvalue{/sea/causal.diff/theme_nameForest:fml.user.all}{0.052891642027139}
\pgfkeyssetvalue{/sea/causal.diff/theme_nameGhost house:fml.all.both}{-0.0438229976181758}
\pgfkeyssetvalue{/sea/causal.diff/theme_nameGhost house:fml.user.all}{-0.0394645709617482}
\pgfkeyssetvalue{/sea/causal.diff/theme_nameOverworld:fml.all.both}{0.0986056265299267}
\pgfkeyssetvalue{/sea/causal.diff/theme_nameOverworld:fml.user.all}{0.103421190808892}
\pgfkeyssetvalue{/sea/causal.diff/theme_nameSky:fml.all.both}{0.0201754674897499}
\pgfkeyssetvalue{/sea/causal.diff/theme_nameSky:fml.user.all}{0.0275956478208172}
\pgfkeyssetvalue{/sea/causal.diff/theme_nameSnow:fml.all.both}{0.0362260719018768}
\pgfkeyssetvalue{/sea/causal.diff/theme_nameSnow:fml.user.all}{0.039953396442487}
\pgfkeyssetvalue{/sea/causal.diff/theme_nameUnderground:fml.all.both}{0.0319106566669227}
\pgfkeyssetvalue{/sea/causal.diff/theme_nameUnderground:fml.user.all}{0.0360119256788045}
\pgfkeyssetvalue{/sea/causal.diff/theme_nameUnderwater:fml.all.both}{0.280308302727238}
\pgfkeyssetvalue{/sea/causal.diff/theme_nameUnderwater:fml.user.all}{0.279627135194563}
\pgfkeyssetvalue{/sea/causal.diff/game_versionV1.0.1:fml.all.both}{-0.179409318677557}
\pgfkeyssetvalue{/sea/causal.diff/game_versionV1.0.1:fml.user.all}{-0.177451941691362}
\pgfkeyssetvalue{/sea/causal.diff/game_versionV1.1.0:fml.all.both}{-0.0615599035637824}
\pgfkeyssetvalue{/sea/causal.diff/game_versionV1.1.0:fml.user.all}{-0.0473000699962012}
\pgfkeyssetvalue{/sea/causal.diff/game_versionV2.0.0:fml.all.both}{-0.0216708480577948}
\pgfkeyssetvalue{/sea/causal.diff/game_versionV2.0.0:fml.user.all}{-0.012164549751524}
\pgfkeyssetvalue{/sea/causal.diff/game_versionV3.0.0:fml.all.both}{0.0271476402675078}
\pgfkeyssetvalue{/sea/causal.diff/game_versionV3.0.0:fml.user.all}{0.0306399572784977}
\pgfkeyssetvalue{/sea/causal.diff/game_versionV3.0.1:fml.all.both}{0.0692702218913085}
\pgfkeyssetvalue{/sea/causal.diff/game_versionV3.0.1:fml.user.all}{0.0734808531330322}
\pgfkeyssetvalue{/sea/causal.diff/upload_attempts:fml.all.both}{-0.172378921314361}
\pgfkeyssetvalue{/sea/causal.diff/upload_attempts:fml.user.all}{-0.173872812881465}
\pgfkeyssetvalue{/sea/causal.insignificant.vars/::N_vars}{1}
\pgfkeyssetvalue{/sea/factor/difficulty_nameEasy:Variable}{difficulty\_name}
\pgfkeyssetvalue{/sea/factor/difficulty_nameNormal:Variable}{difficulty\_name}
\pgfkeyssetvalue{/sea/factor/difficulty_nameExpert:Variable}{difficulty\_name}
\pgfkeyssetvalue{/sea/factor/difficulty_nameSuper expert:Variable}{difficulty\_name}
\pgfkeyssetvalue{/sea/factor/gamestyle_nameSMB1:Variable}{gamestyle\_name}
\pgfkeyssetvalue{/sea/factor/gamestyle_nameSMB3:Variable}{gamestyle\_name}
\pgfkeyssetvalue{/sea/factor/gamestyle_nameSMW:Variable}{gamestyle\_name}
\pgfkeyssetvalue{/sea/factor/gamestyle_nameNSMBU:Variable}{gamestyle\_name}
\pgfkeyssetvalue{/sea/factor/gamestyle_nameSM3DW:Variable}{gamestyle\_name}
\pgfkeyssetvalue{/sea/factor/theme_nameAirship:Variable}{theme\_name}
\pgfkeyssetvalue{/sea/factor/theme_nameCastle:Variable}{theme\_name}
\pgfkeyssetvalue{/sea/factor/theme_nameDesert:Variable}{theme\_name}
\pgfkeyssetvalue{/sea/factor/theme_nameForest:Variable}{theme\_name}
\pgfkeyssetvalue{/sea/factor/theme_nameGhost house:Variable}{theme\_name}
\pgfkeyssetvalue{/sea/factor/theme_nameOverworld:Variable}{theme\_name}
\pgfkeyssetvalue{/sea/factor/theme_nameSky:Variable}{theme\_name}
\pgfkeyssetvalue{/sea/factor/theme_nameSnow:Variable}{theme\_name}
\pgfkeyssetvalue{/sea/factor/theme_nameUnderground:Variable}{theme\_name}
\pgfkeyssetvalue{/sea/factor/theme_nameUnderwater:Variable}{theme\_name}
\pgfkeyssetvalue{/sea/factor/game_versionV1.0.0:Variable}{game\_version}
\pgfkeyssetvalue{/sea/factor/game_versionV1.0.1:Variable}{game\_version}
\pgfkeyssetvalue{/sea/factor/game_versionV1.1.0:Variable}{game\_version}
\pgfkeyssetvalue{/sea/factor/game_versionV2.0.0:Variable}{game\_version}
\pgfkeyssetvalue{/sea/factor/game_versionV3.0.0:Variable}{game\_version}
\pgfkeyssetvalue{/sea/factor/game_versionV3.0.1:Variable}{game\_version}
\pgfkeyssetvalue{/sea/insignificant.vars/data_99_9_makers:fml.all.both:data}{data\_99\_9\_makers}
\pgfkeyssetvalue{/sea/insignificant.vars/data_99_9_makers:fml.all.both:model}{fml.all.both}
\pgfkeyssetvalue{/sea/insignificant.vars/data_99_9_makers:fml.user.all:data}{data\_99\_9\_makers}
\pgfkeyssetvalue{/sea/insignificant.vars/data_99_9_makers:fml.user.all:model}{fml.user.all}
\pgfkeyssetvalue{/sea/insignificant.vars/data_99_9_players:fml.all.both:data}{data\_99\_9\_players}
\pgfkeyssetvalue{/sea/insignificant.vars/data_99_9_players:fml.all.both:model}{fml.all.both}
\pgfkeyssetvalue{/sea/insignificant.vars/data_99_9_players:fml.user.all:data}{data\_99\_9\_players}
\pgfkeyssetvalue{/sea/insignificant.vars/data_99_9_players:fml.user.all:model}{fml.user.all}
\pgfkeyssetvalue{/sea/causal.insignificant.vars/::model.all}{fml.all.both} \pgfkeyssetvalue{/sea/avg_S_i_comment:normal:neg}{0.22729908116957206}
\pgfkeyssetvalue{/sea/avg_S_i_comment:normal:neu}{0.3812437125480982}
\pgfkeyssetvalue{/sea/avg_S_i_comment:normal:pos}{0.3914572063869114}
\pgfkeyssetvalue{/sea/avg_S_i_comment:superexpert:neg}{0.2927489672055229}
\pgfkeyssetvalue{/sea/avg_S_i_comment:superexpert:neu}{0.34576174191544523}
\pgfkeyssetvalue{/sea/avg_S_i_comment:superexpert:pos}{0.3614892909555892}
\pgfkeyssetvalue{/sea/avg_S_i_comment:easy:neg}{0.2304440952549333}
\pgfkeyssetvalue{/sea/avg_S_i_comment:easy:neu}{0.40139909298002224}
\pgfkeyssetvalue{/sea/avg_S_i_comment:easy:pos}{0.3681568118484991}
\pgfkeyssetvalue{/sea/avg_S_i_comment:expert:neg}{0.24516057726171112}
\pgfkeyssetvalue{/sea/avg_S_i_comment:expert:neu}{0.3574641342448585}
\pgfkeyssetvalue{/sea/avg_S_i_comment:expert:pos}{0.39737528859551186}
\pgfkeyssetvalue{/sea/avg_S_i_level:superexpert:neg}{0.2736746457179461}
\pgfkeyssetvalue{/sea/avg_S_i_level:superexpert:neu}{0.34132598613849124}
\pgfkeyssetvalue{/sea/avg_S_i_level:superexpert:pos}{0.38499936817750385}
\pgfkeyssetvalue{/sea/avg_S_i_level:easy:neg}{0.22134664132376075}
\pgfkeyssetvalue{/sea/avg_S_i_level:easy:neu}{0.3939561202230435}
\pgfkeyssetvalue{/sea/avg_S_i_level:easy:pos}{0.3846972384984446}
\pgfkeyssetvalue{/sea/avg_S_i_level:expert:neg}{0.2294684102281084}
\pgfkeyssetvalue{/sea/avg_S_i_level:expert:neu}{0.34898462474112524}
\pgfkeyssetvalue{/sea/avg_S_i_level:expert:pos}{0.4215469651506307}
\pgfkeyssetvalue{/sea/avg_S_i_level:normal:neg}{0.20940032721228405}
\pgfkeyssetvalue{/sea/avg_S_i_level:normal:neu}{0.3657862339996966}
\pgfkeyssetvalue{/sea/avg_S_i_level:normal:pos}{0.4248134388853808}
\pgfkeyssetvalue{/sea/sent:comments:N.comments}{666}
\pgfkeyssetvalue{/sea/sent:comments:N.levels}{666}

\pgfkeyssetvalue{/sea/topics:desc:top_n_superexpert:15:Easy_p}{0.01897486145415264}
\pgfkeyssetvalue{/sea/topics:desc:top_n_superexpert:15:Normal_p}{0.022792859445652264}
\pgfkeyssetvalue{/sea/topics:desc:top_n_superexpert:15:Expert_p}{0.03743632867642162}
\pgfkeyssetvalue{/sea/topics:desc:top_n_superexpert:15:Super_Expert_p}{0.044510495282172526}
\pgfkeyssetvalue{/sea/topics:desc:top_n_superexpert:13:Easy_p}{0.00927723006718165}
\pgfkeyssetvalue{/sea/topics:desc:top_n_superexpert:13:Normal_p}{0.015936257450132676}
\pgfkeyssetvalue{/sea/topics:desc:top_n_superexpert:13:Expert_p}{0.025467446470603845}
\pgfkeyssetvalue{/sea/topics:desc:top_n_superexpert:13:Super_Expert_p}{0.03150891532517278}
\pgfkeyssetvalue{/sea/topics:desc:top_n_superexpert:32:Easy_p}{0.0021759813794909543}
\pgfkeyssetvalue{/sea/topics:desc:top_n_superexpert:32:Normal_p}{0.004934653759992064}
\pgfkeyssetvalue{/sea/topics:desc:top_n_superexpert:32:Expert_p}{0.01562712361834514}
\pgfkeyssetvalue{/sea/topics:desc:top_n_superexpert:32:Super_Expert_p}{0.028026038019445466}
\pgfkeyssetvalue{/sea/topics:desc:top_n_superexpert:56:Easy_p}{0.0007040813933942502}
\pgfkeyssetvalue{/sea/topics:desc:top_n_superexpert:56:Normal_p}{0.001476593563746683}
\pgfkeyssetvalue{/sea/topics:desc:top_n_superexpert:56:Expert_p}{0.0060618727965133}
\pgfkeyssetvalue{/sea/topics:desc:top_n_superexpert:56:Super_Expert_p}{0.023710633079849854}
\pgfkeyssetvalue{/sea/topics:desc:top_n_superexpert:25:Easy_p}{0.0036181593555475633}
\pgfkeyssetvalue{/sea/topics:desc:top_n_superexpert:25:Normal_p}{0.005904720966181978}
\pgfkeyssetvalue{/sea/topics:desc:top_n_superexpert:25:Expert_p}{0.010717521059525479}
\pgfkeyssetvalue{/sea/topics:desc:top_n_superexpert:25:Super_Expert_p}{0.014540525743251348}

\pgfkeyssetvalue{/sea/topics:desc:top_n_easy:0:Easy_p}{0.09514609854035208}
\pgfkeyssetvalue{/sea/topics:desc:top_n_easy:0:Normal_p}{0.07571318745815113}
\pgfkeyssetvalue{/sea/topics:desc:top_n_easy:0:Expert_p}{0.05896558838069753}
\pgfkeyssetvalue{/sea/topics:desc:top_n_easy:0:Super_Expert_p}{0.04750021274566998}
\pgfkeyssetvalue{/sea/topics:desc:top_n_easy:5:Easy_p}{0.03856496851613525}
\pgfkeyssetvalue{/sea/topics:desc:top_n_easy:5:Normal_p}{0.03463640045961429}
\pgfkeyssetvalue{/sea/topics:desc:top_n_easy:5:Expert_p}{0.03572416773112684}
\pgfkeyssetvalue{/sea/topics:desc:top_n_easy:5:Super_Expert_p}{0.03689471293688477}
\pgfkeyssetvalue{/sea/topics:desc:top_n_easy:30:Easy_p}{0.023573186651911337}
\pgfkeyssetvalue{/sea/topics:desc:top_n_easy:30:Normal_p}{0.01807023997487001}
\pgfkeyssetvalue{/sea/topics:desc:top_n_easy:30:Expert_p}{0.009781301491399396}
\pgfkeyssetvalue{/sea/topics:desc:top_n_easy:30:Super_Expert_p}{0.006546414953508662}
\pgfkeyssetvalue{/sea/topics:desc:top_n_easy:16:Easy_p}{0.01887347637546032}
\pgfkeyssetvalue{/sea/topics:desc:top_n_easy:16:Normal_p}{0.01782265997635797}
\pgfkeyssetvalue{/sea/topics:desc:top_n_easy:16:Expert_p}{0.017664172788553537}
\pgfkeyssetvalue{/sea/topics:desc:top_n_easy:16:Super_Expert_p}{0.016101002416585755}
\pgfkeyssetvalue{/sea/topics:desc:top_n_easy:40:Easy_p}{0.01851417029853117}
\pgfkeyssetvalue{/sea/topics:desc:top_n_easy:40:Normal_p}{0.010128253879028859}
\pgfkeyssetvalue{/sea/topics:desc:top_n_easy:40:Expert_p}{0.007291033247436497}
\pgfkeyssetvalue{/sea/topics:desc:top_n_easy:40:Super_Expert_p}{0.005687230030871191}

\pgfkeyssetvalue{/sea/sent:titledesc:N.levels}{10689031}
\pgfkeyssetvalue{/sea/sent:comments:N.comments}{1996367}
\pgfkeyssetvalue{/sea/sent:comments:N.levels}{100148}

\pgfkeyssetvalue{/sea/topics:comm:top_n_superexpert:0:Easy_p}{0.39296127679165177}
\pgfkeyssetvalue{/sea/topics:comm:top_n_superexpert:0:Normal_p}{0.4545666491810923}
\pgfkeyssetvalue{/sea/topics:comm:top_n_superexpert:0:Expert_p}{0.4887853370611992}
\pgfkeyssetvalue{/sea/topics:comm:top_n_superexpert:0:Super_Expert_p}{0.5566206183629357}
\pgfkeyssetvalue{/sea/topics:comm:top_n_superexpert:5:Easy_p}{0.09821474244206865}
\pgfkeyssetvalue{/sea/topics:comm:top_n_superexpert:5:Normal_p}{0.12155388471177946}
\pgfkeyssetvalue{/sea/topics:comm:top_n_superexpert:5:Expert_p}{0.17639018328673503}
\pgfkeyssetvalue{/sea/topics:comm:top_n_superexpert:5:Super_Expert_p}{0.2224414998590358}
\pgfkeyssetvalue{/sea/topics:comm:top_n_superexpert:10:Easy_p}{0.07005473425750679}
\pgfkeyssetvalue{/sea/topics:comm:top_n_superexpert:10:Normal_p}{0.07920965203706942}
\pgfkeyssetvalue{/sea/topics:comm:top_n_superexpert:10:Expert_p}{0.11102826964895932}
\pgfkeyssetvalue{/sea/topics:comm:top_n_superexpert:10:Super_Expert_p}{0.1380509350624941}
\pgfkeyssetvalue{/sea/topics:comm:top_n_superexpert:50:Easy_p}{0.014860095145531741}
\pgfkeyssetvalue{/sea/topics:comm:top_n_superexpert:50:Normal_p}{0.026228361601678615}
\pgfkeyssetvalue{/sea/topics:comm:top_n_superexpert:50:Expert_p}{0.04137931034482759}
\pgfkeyssetvalue{/sea/topics:comm:top_n_superexpert:50:Super_Expert_p}{0.08852551451931209}
\pgfkeyssetvalue{/sea/topics:comm:top_n_superexpert:38:Easy_p}{0.037776868382014425}
\pgfkeyssetvalue{/sea/topics:comm:top_n_superexpert:38:Normal_p}{0.03820598006644518}
\pgfkeyssetvalue{/sea/topics:comm:top_n_superexpert:38:Expert_p}{0.04628766697732215}
\pgfkeyssetvalue{/sea/topics:comm:top_n_superexpert:38:Super_Expert_p}{0.08814960999906024}

\pgfkeyssetvalue{/sea/topics:comm:top_n_easy:-1:Easy_p}{0.9974679011714155}
\pgfkeyssetvalue{/sea/topics:comm:top_n_easy:-1:Normal_p}{0.9970565949758117}
\pgfkeyssetvalue{/sea/topics:comm:top_n_easy:-1:Expert_p}{0.9962100031065549}
\pgfkeyssetvalue{/sea/topics:comm:top_n_easy:-1:Super_Expert_p}{0.997368668358237}
\pgfkeyssetvalue{/sea/topics:comm:top_n_easy:1:Easy_p}{0.2411120773441097}
\pgfkeyssetvalue{/sea/topics:comm:top_n_easy:1:Normal_p}{0.21489770938975347}
\pgfkeyssetvalue{/sea/topics:comm:top_n_easy:1:Expert_p}{0.1636533084808947}
\pgfkeyssetvalue{/sea/topics:comm:top_n_easy:1:Super_Expert_p}{0.1242364439432384}
\pgfkeyssetvalue{/sea/topics:comm:top_n_easy:2:Easy_p}{0.169471584224257}
\pgfkeyssetvalue{/sea/topics:comm:top_n_easy:2:Normal_p}{0.14373142157719881}
\pgfkeyssetvalue{/sea/topics:comm:top_n_easy:2:Expert_p}{0.14290152221186705}
\pgfkeyssetvalue{/sea/topics:comm:top_n_easy:2:Super_Expert_p}{0.1490461422798609}
\pgfkeyssetvalue{/sea/topics:comm:top_n_easy:3:Easy_p}{0.16146605964499464}
\pgfkeyssetvalue{/sea/topics:comm:top_n_easy:3:Normal_p}{0.14999708573759982}
\pgfkeyssetvalue{/sea/topics:comm:top_n_easy:3:Expert_p}{0.13793103448275865}
\pgfkeyssetvalue{/sea/topics:comm:top_n_easy:3:Super_Expert_p}{0.11737618644864203}
\pgfkeyssetvalue{/sea/topics:comm:top_n_easy:4:Easy_p}{0.16115913857486316}
\pgfkeyssetvalue{/sea/topics:comm:top_n_easy:4:Normal_p}{0.14309028384915778}
\pgfkeyssetvalue{/sea/topics:comm:top_n_easy:4:Expert_p}{0.12581547064305687}
\pgfkeyssetvalue{/sea/topics:comm:top_n_easy:4:Super_Expert_p}{0.09444601071327882}
\pgfkeyssetvalue{/sea/topics:comm:top_n_easy:8:Easy_p}{0.12420072637986598}
\pgfkeyssetvalue{/sea/topics:comm:top_n_easy:8:Normal_p}{0.09177012298187329}
\pgfkeyssetvalue{/sea/topics:comm:top_n_easy:8:Expert_p}{0.09388008698353527}
\pgfkeyssetvalue{/sea/topics:comm:top_n_easy:8:Super_Expert_p}{0.09801710365567146}

\pgfkeyssetvalue{/sea/avg_SR_i:superexpert:neg}{0.26481370814623006}
\pgfkeyssetvalue{/sea/avg_SR_i:superexpert:neu}{0.36101630209098995}
\pgfkeyssetvalue{/sea/avg_SR_i:superexpert:pos}{0.3741699897627799}
\pgfkeyssetvalue{/sea/avg_SR_i:easy:neg}{0.1977866141420175}
\pgfkeyssetvalue{/sea/avg_SR_i:easy:neu}{0.44031133739031253}
\pgfkeyssetvalue{/sea/avg_SR_i:easy:pos}{0.3619020484676698}
\pgfkeyssetvalue{/sea/avg_SR_i:expert:neg}{0.2124844869567172}
\pgfkeyssetvalue{/sea/avg_SR_i:expert:neu}{0.37798632169603835}
\pgfkeyssetvalue{/sea/avg_SR_i:expert:pos}{0.4095291913472441}
\pgfkeyssetvalue{/sea/avg_SR_i:normal:neg}{0.18645783660401782}
\pgfkeyssetvalue{/sea/avg_SR_i:normal:neu}{0.4048855193474178}
\pgfkeyssetvalue{/sea/avg_SR_i:normal:pos}{0.40865664404856433}

\pgfkeyssetvalue{/sea/topics:comm:0:sent_pos:avg}{0.665554}
\pgfkeyssetvalue{/sea/topics:comm:38:sent_neg:avg}{0.459021}
\pgfkeyssetvalue{/sea/topics:comm:3:sent_pos:avg}{0.624591}
\pgfkeyssetvalue{/sea/topics:comm:4:sent_pos:avg}{0.857831}
\pgfkeyssetvalue{/sea/topics:comm:8:sent_neg:avg}{0.341711}

\pgfkeyssetvalue{/sea/topics:comm:2:sent_neg:avg}{0.202001}

\pgfkeyssetvalue{/sea/d_S_i_comment:easy:neg}{-0.02512335809691173}
\pgfkeyssetvalue{/sea/d_S_i_comment:normal:neg}{-0.0942340022722234}
\pgfkeyssetvalue{/sea/d_S_i_comment:expert:neg}{0.030697667728079983}
\pgfkeyssetvalue{/sea/d_S_i_comment:superexpert:neg}{0.24286592404094834}
\pgfkeyssetvalue{/sea/d_S_i_comment:easy:pos}{-0.08635772710661452}
\pgfkeyssetvalue{/sea/d_S_i_comment:normal:pos}{0.07954936541916316}
\pgfkeyssetvalue{/sea/d_S_i_comment:expert:pos}{0.05780053929567147}
\pgfkeyssetvalue{/sea/d_S_i_comment:superexpert:pos}{-0.054353446193786514}

\pgfkeyssetvalue{/sea/d_SR_i_comment:easy:neg}{-0.04134384967419905}
\pgfkeyssetvalue{/sea/d_SR_i_comment:normal:neg}{-0.086964355370354}
\pgfkeyssetvalue{/sea/d_SR_i_comment:expert:neg}{0.043784077907293266}
\pgfkeyssetvalue{/sea/d_SR_i_comment:superexpert:neg}{0.24768632509619698}
\pgfkeyssetvalue{/sea/d_SR_i_comment:easy:pos}{-0.09412821072599066}
\pgfkeyssetvalue{/sea/d_SR_i_comment:normal:pos}{0.07164135233539722}
\pgfkeyssetvalue{/sea/d_SR_i_comment:expert:pos}{0.06456537927461241}
\pgfkeyssetvalue{/sea/d_SR_i_comment:superexpert:pos}{-0.025731977244347913}

\pgfkeyssetvalue{/sea/cliffsdelta:S_i:comment}{0.09335058085813255}
\pgfkeyssetvalue{/sea/cliffsdelta:S_i:level}{0.19272765469357053}
\pgfkeyssetvalue{/sea/cliffsdelta:SR_i}{0.24768632509619698} 

\usepackage{mathtools}

\DeclareDocumentCommand{\spellout}{m O{13}}{\IfStrEq{#1}{ }{\NAN}{\IfInteger{#1}{\expandafter\ifnum#1<#2{\numberstringnum{#1}}\else#1\fi }{\NAN}}}

\DeclareDocumentCommand{\n}{t. t: o m o O{} t| t!}{\begingroup \pgfkeys{/pgf/fpu=true}\IfBooleanTF{#2}{\pgfkeyssetvalue{/tmp/value}{#4}\pgfkeyssetvalue{/tmp/found}{found}}{\pgfkeysifdefined{\keyfamily#4}{\pgfkeyssetvalue{/tmp/value}{\pgfkeysvalueof{\keyfamily#4}}\pgfkeyssetvalue{/tmp/found}{found}}{}}\pgfkeysifdefined{/tmp/found}{\IfNoValueF{#5}{\pgfkeyssetvalue{/tmp/multiplier}{#5}\pgfmathparse{\pgfkeysvalueof{/tmp/multiplier} * \pgfkeysvalueof{/tmp/value}}\pgfkeyslet{/tmp/value}\pgfmathresult }\IfBooleanTF{#1}{\IfBooleanTF{#8}{\spellout{\pgfkeysvalueof{/tmp/value}}}{\pgfkeysvalueof{/tmp/value}}}{\IfNoValueTF{#3}{\pgfmathprintnumber [set thousands separator={\,},int detect,#6]{\pgfkeysvalueof{/tmp/value}}}{{\pgfmathprintnumber [precision=#3,fixed,zerofill,set thousands separator={\,},#6]{\pgfkeysvalueof{/tmp/value}}}}}\IfBooleanT{#7}{{\smaller[1.2]\%}}}{\NAN }\pgfkeys{/pgf/fpu=false}\endgroup }

\usepackage{tcolorbox}

\DeclareRobustCommand{\dist}[1]{\ensuremath{\textsf{#1}}}

\DeclareDocumentCommand{\dcoef}{s o}{\ensuremath{\delta\IfNoValueF{#2}{\IfBooleanTF{#1}{#2}{(\var{#2})}}}}

\DeclareDocumentCommand{\ddiff}{s m m o}{\ensuremath{\tau\IfBooleanF{#1}{^{#3}_{#2}}\IfNoValueF{#4}{(\var{#4})}}}

\DeclareDocumentCommand{\Dcoef}{s o}{\ensuremath{\Delta\IfNoValueF{#2}{\IfBooleanTF{#1}{#2}{(\var{#2})}}}}

\DeclareDocumentCommand{\mad}{s o}{\ensuremath{\mathsf{md}\IfNoValueF{#2}{\IfBooleanTF{#1}{#2}{(\var{#2})}}}}

\DeclareDocumentCommand{\SR}{}{\ensuremath{\mathit{D}}}

\DeclareDocumentCommand{\bracketcol}{s m m O{black} O{black} O{black}}
{\IfBooleanTF{#1}{\begingroup
    \color{#4}
    \overbracket{\color{#6}#2}^{\color{#5}#3}
    \endgroup
  }{\begingroup
    \color{#4}
    \underbracket{\color{#6}#2}_{\color{#5}#3}
    \endgroup
  }}

\newcommand{\smm}{{\smaller \textsc{smm2}}\xspace}
\newcommand{\bertopic}{{\smaller \textsc{BERTopic}}\xspace}

\DeclareRobustCommand{\var}[1]{\ensuremath{\textnormal{\textsl{#1}}}}

\DeclareRobustCommand{\coef}[1]{\ensuremath{\beta_{\var{#1}}}}

\DeclareDocumentCommand{\vars}{o}{\ensuremath{\mathcal{\IfNoValueTF{#1}{V}{#1}}}}

\DeclareMathOperator*{\argmax}{argmax}

\begin{document}

\title{What Makes a Level Hard in Super Mario Maker 2?\ifblind\else\\
  \thanks{Work partially supported by SNF grant 200021-207919 (LastMile).}\fi }

\ifblind
\author{}
\else
\author{\IEEEauthorblockN{Carlo A. Furia}
\IEEEauthorblockA{\textit{Software Institute} --
\textit{USI Università della Svizzera italiana}\\
Lugano, Switzerland \\
\url{bugcounting.net}}
\and
\IEEEauthorblockN{Andrea Mocci}
\IEEEauthorblockA{\textit{Software Institute} --
\textit{USI Università della Svizzera italiana}\\
Lugano, Switzerland \\
\url{andrea.mocci@usi.ch}}
}
\fi

\maketitle

\begin{abstract}
  Games like Super Mario Maker~2 (\smm)
  lower the barrier for casual users
  to become level \emph{designers}.
  In this paper, we set out to analyze
  a vast amount of data about \smm
  user-written levels, in order to understand
  what factors affect a level's \emph{difficulty}
  as experienced by other users.
  To this end, we perform two kinds of analyses:
  one based on regression models and one using natural language processing
  techniques.
  The main results shed light on which level characteristics
  (e.g., its style, popularity, timing)
  and which topics and sentiments
  have a consistent association with easier or harder levels.
  While none of our findings are startling,
  they help distill some key differences between easy and hard \smm levels,
  which, in turn, can pave the way for a better understanding of
  end-user level design.
\end{abstract}


\pagestyle{plain}

\section{Introduction}
\label{sec:intro}

Super Mario Maker~2 (\smm) is a popular game for the Nintendo Switch
that sold over \spellout{8} million copies
since its release in 2019.\footnote{Source: \url{https://en.wikipedia.org/wiki/Super_Mario_Maker_2}}
The game's key innovation is combining
a 2D platformer with a level editor:
\smm users can just play the levels provided
with the game, but can also create their own levels
and share them with other users by uploading them to Nintendo's servers.
Thus, every \smm user can be both a player and a \emph{maker};
and we can say that \smm makers engage in a form of \emph{end-user programming}.

Starting with its predecessor for the Wii~U,
a vibrant community of passionate gamers and content creators
coalesced around the game.
As part of their efforts,
a large dump of \smm level data
has been collected, reverse engineered, and made publicly available~\cite{mariomaker2:dataset}.
Broadly speaking, our goal is analyzing these \smm data
with techniques commonly used in empirical software engineering research,
in order to shed some light on this peculiar form of end-user programming.

In this paper,
we focus on a fundamental question:
\emph{what denotes the difficulty of a level in \smm?}
Even casual players are well aware of the broad
range of level difficulties one encounters in \smm ---from facile ``little Timmy'' levels
to unforgiving kaizos.\footnote{An overview of \smm's lingo, which we'll occasionally use in the paper: \url{https://supermariomaker2.fandom.com/wiki/Terminology}}
Finding out what characteristics of a level
are associated with its difficulty 
is thus a fundamental way of understanding
how \smm users harness the game's features
to create a broad variety of challenges and experiences.
Our analysis considers both intrinsic level characteristics
(e.g., its visual style)
and community engagement
(e.g., how many likes a level received).

After an overview of the data,
we tackle the question in two 
ways.
First, we perform a regression analysis of
the level data to determine
which variables contribute the most to a level's
clear rate (a fundamental measure of its difficulty).
Second,
we use NLP machine learning techniques to perform
a topic classification
of the levels' titles, descriptions, and user comments,
which is the basis for a qualitative analysis
of how certain topics and sentiments are linked to a level's difficulty.
These two analyses are complementary in the data they use
and the insights they provide.

\ifextended
In summary, the paper makes the following contributions:
\begin{enumerate}
\item Regression analyses of the metadata of over 26 million \smm levels.
\item NLP topic and sentiment analyses of the title, description,
  and user comments of the over 10 million \smm levels with title, description, or comments in English.
\item For reproducibility, the detailed analysis results and all the analysis scripts are available in a replication package.\footnote{\ifblind
      To be published after double-anonymous review.
    \else
      \url{https://dx.doi.org/10.6084/m9.figshare.28525223}
    \fi }  
\end{enumerate}
\else
  \titledsubpar{Data availability.}
  For reproducibility, the detailed results and the analysis scripts are available in a replication package.\footnote{\ifblind
      To be published after double-anonymous review.
    \else
      \url{https://dx.doi.org/10.6084/m9.figshare.28525223}
    \fi }
\fi

\section{Data Overview}
\label{sec:data}

The dump of \smm data we analyzed
covers a whopping \n{N.levels} levels---a comprehensive snapshot
taken in February 2022~\cite{mariomaker2:dataset}.
Let us give an overview of the level and user data,
which we analyze in~\autoref{sec:regression}.

\begin{table*}[!htb]
  \centering
  \scriptsize
  \begin{subtable}{\linewidth}
    \centering
    \setlength{\tabcolsep}{6pt}
    \renewcommand{\arraystretch}{0.9}
    \begin{tabular}{lrrrrrrrrrrrrrrr}
\toprule
\multicolumn{1}{c}{ } & \multicolumn{7}{c}{\textcolor[HTML]{1B9E77}{plays}} & \multicolumn{3}{c}{\textcolor[HTML]{D95F02}{timing}} & \multicolumn{5}{c}{\textcolor[HTML]{66A61E}{user}} \\
\cmidrule(l{3pt}r{3pt}){2-8} \cmidrule(l{3pt}r{3pt}){9-11} \cmidrule(l{3pt}r{3pt}){12-16}
\rotatebox{-90}{ } & \rotatebox{-90}{\textcolor[HTML]{1B9E77}{attempts}} & \rotatebox{-90}{\textcolor[HTML]{1B9E77}{boos}} & \rotatebox{-90}{\textcolor[HTML]{1B9E77}{clear rate}} & \rotatebox{-90}{\textcolor[HTML]{1B9E77}{clears}} & \rotatebox{-90}{\textcolor[HTML]{1B9E77}{comments}} & \rotatebox{-90}{\textcolor[HTML]{1B9E77}{likes}} & \rotatebox{-90}{\textcolor[HTML]{1B9E77}{players}} & \rotatebox{-90}{\textcolor[HTML]{D95F02}{autoscroll: speed}} & \rotatebox{-90}{\textcolor[HTML]{D95F02}{timer}} & \rotatebox{-90}{\textcolor[HTML]{D95F02}{world record}} & \rotatebox{-90}{\textcolor[HTML]{66A61E}{cleared}} & \rotatebox{-90}{\textcolor[HTML]{66A61E}{first clears}} & \rotatebox{-90}{\textcolor[HTML]{66A61E}{maker points}} & \rotatebox{-90}{\textcolor[HTML]{66A61E}{records}} & \rotatebox{-90}{\textcolor[HTML]{66A61E}{uploaded}}\\
\midrule
min & \numprint{0} & \numprint{0} & \numprint{0.00} & \numprint{0} & \numprint{0} & \numprint{0} & \numprint{0} & \numprint{0} & \numprint{10} & \numprint{-0} & \numprint{0} & \numprint{0} & \numprint{0} & \numprint{0} & \numprint{0}\\
25\% & \numprint{22} & \numprint{0} & \numprint{0.10} & \numprint{4} & \numprint{0} & \numprint{0} & \numprint{10} & \numprint{0} & \numprint{300} & \numprint{10} & \numprint{80} & \numprint{1} & \numprint{780} & \numprint{0} & \numprint{10}\\
50\% & \numprint{54} & \numprint{1} & \numprint{0.25} & \numprint{10} & \numprint{0} & \numprint{1} & \numprint{21} & \numprint{0} & \numprint{300} & \numprint{21} & \numprint{219} & \numprint{4} & \numprint{1599} & \numprint{2} & \numprint{25}\\
75\% & \numprint{127} & \numprint{2} & \numprint{0.50} & \numprint{28} & \numprint{1} & \numprint{4} & \numprint{48} & \numprint{0} & \numprint{300} & \numprint{44} & \numprint{534} & \numprint{11} & \numprint{2441} & \numprint{7} & \numprint{52}\\
99\% & \numprint{1431} & \numprint{15} & \numprint{1.00} & \numprint{247} & \numprint{13} & \numprint{49} & \numprint{325} & \numprint{0} & \numprint{500} & \numprint{258} & \numprint{4502} & \numprint{215} & \numprint{7687} & \numprint{268} & \numprint{100}\\
99.99\% & \numprint{142163} & \numprint{576} & \numprint{1.00} & \numprint{15557} & \numprint{906} & \numprint{4590} & \numprint{18171} & \numprint{2} & \numprint{500} & \numprint{2628} & \numprint{57813} & \numprint{7803} & \numprint{22633} & \numprint{10674} & \numprint{100}\\
max & \numprint{67895905} & \numprint{37886} & \numprint{1.00} & \numprint{1804430} & \numprint{120515} & \numprint{564890} & \numprint{1302862} & \numprint{2} & \numprint{500} & \numprint{6000} & \numprint{513124} & \numprint{67102} & \numprint{29100} & \numprint{105379} & \numprint{100}\\
\bottomrule
\end{tabular}
     \caption{An overview of the main \emph{numeric} variables in \smm's data.
      Variables are grouped according to whether they refer
      to a level's \textcolor{plays}{plays} statistics,
      \textcolor{timing}{timing}, or the \textcolor{user}{user} who created it.
      For each variable, the table reports the minimum, maximum as well as the 25\%, 50\%, 75\%, 99\%, and 99.99\% percentiles in the data.}
    \label{tab:numeric}
  \end{subtable}
  \begin{subtable}{\linewidth}
    \centering
    \setlength{\tabcolsep}{6pt}
    \renewcommand{\arraystretch}{1.2}
    \begin{tabular}{ccccccccccccccccccccccccc}
\toprule
\multicolumn{4}{c}{\textcolor[HTML]{E7298A}{difficulty}} & \multicolumn{5}{c}{\textcolor[HTML]{7570B3}{style}} & \multicolumn{10}{c}{\textcolor[HTML]{7570B3}{theme}} & \multicolumn{6}{c}{\textcolor[HTML]{7570B3}{version}} \\
\cmidrule(l{3pt}r{3pt}){1-4} \cmidrule(l{3pt}r{3pt}){5-9} \cmidrule(l{3pt}r{3pt}){10-19} \cmidrule(l{3pt}r{3pt}){20-25}
\rotatebox{-90}{easy} & \rotatebox{-90}{normal} & \rotatebox{-90}{expert} & \rotatebox{-90}{super expert} & \rotatebox{-90}{SMB1} & \rotatebox{-90}{SMB3} & \rotatebox{-90}{SMW} & \rotatebox{-90}{NSMBU} & \rotatebox{-90}{SM3DW} & \rotatebox{-90}{airship} & \rotatebox{-90}{castle} & \rotatebox{-90}{desert} & \rotatebox{-90}{forest} & \rotatebox{-90}{ghost} & \rotatebox{-90}{overworld} & \rotatebox{-90}{sky} & \rotatebox{-90}{snow} & \rotatebox{-90}{underground} & \rotatebox{-90}{water} & \rotatebox{-90}{1.0.0} & \rotatebox{-90}{1.0.1} & \rotatebox{-90}{1.1.0} & \rotatebox{-90}{2.0.0} & \rotatebox{-90}{3.0.0} & \rotatebox{-90}{3.0.1}\\
33 & 44 & 16 & 8 & 14 & 8 & 17 & 33 & 28 & 6 & 15 & 7 & 8 & 5 & 31 & 10 & 9 & 7 & 2 & 0 & 24 & 5 & 15 & 14 & 42\\
\bottomrule
\end{tabular}
     \caption{An overview of the main \emph{nominal} variables in \smm's data.
      Variables are colored according to whether they refer
      to a level's \textcolor{style}{style} or to its \textcolor{difficulty}{difficulty} rating.
      For each discrete value of each variable, the table reports the
      \emph{percentage} of levels with that value.
    }
    \label{tab:factor}
  \end{subtable}
  \caption{An overview of the level and user data in \smm.}
  \label{tab:data-overview}
\end{table*}

\subsection{Level Data}
The level data include
\n{N.raw.vars} variables;
after culling undocumented and redundant
variables, as well as
others that are unsuitable for a regression analysis
(e.g., text such as a level's title or user comments),
we ended up with
a selection of \n{N.all.variables}
variables, which characterize each level
along different dimensions.
Variables are of two main types:
numeric and nominal (possibly ordinal);
\autoref{tab:numeric} and \autoref{tab:factor}
overview several variables of each kind.
Independent of their type, we group and color variables
into categories (\vg{style}, \vg{plays}, \vg{timing}, and \vg{difficulty})
according to what aspect of a level's design they pertain to.
Rather than tediously going through all variables systematically,
let's illustrate those that are most relevant for \autoref{sec:regression}'s
analysis, while explaining how levels are made and played in \smm.

When they create a level in the game's editor,
\smm users can choose a \vg*{style}[style]
among four classic Nintendo games: SMB1 (Super Mario Brothers 1),
SMB3 (Super Mario Brothers 3), SMW (Super Mario World),
NSMBU (New Super Mario Brothers~U), and SM3DW (Super Mario 3D World).
A level's \vg*{style}[theme] denotes
the styling of its graphical elements
to resemble environments such as 
a \emph{castle}, a winter landscape with \emph{snow},
a \emph{forest}, etc.
Throughout its life, \smm went through several \vg*{style}[version] numbers,
which affect some of the features available in the editor.
\autoref{tab:factor} shows that, in the dataset we analyzed,
the most popular game \var{style} is NSMBU,
but all styles are fairly used;
the most popular \var{theme} is \emph{overworld} (possibly simply because it is the default for each style);
and the game \var{version} with more levels is 3.0.1
(probably just because it was the most recent when the data was collected).
Every level also has a \vg*{timing}[timer] (the number of seconds a player has to clear it), and may feature \vg{timing}[autoscroll] with different speeds,
as well as a clear condition (e.g., ``do not take damage'').
\autoref{tab:numeric} shows that the median level\footnote{
  With a little abuse of terminology, ``median level'' denotes
  a hypothetical level all of whose variables are at the level of the median.
}
has a timer of \n{numeric/50p:timer} seconds (the default in the editor)
and no autoscroll.
Before a user is allowed to upload a level,
they must clear it to show that it can be beaten.
The system records the number of attempts (\vg*{timing}[upload attempts])
and the time (\vg*{timing}[upload time]) taken by the maker to clear a level before uploading it,
as well as the timestamp of when the upload finally took place, 
and the maker's user id.

Once a level is uploaded, any \smm user can play it.
For each level,
the system keeps track of
the total number of users who played the level (\vg*{plays}[players])
the total number of \vg*{plays}[attempts],
and of successful \vg*{plays}[clears];
the id of the first user who cleared the level and of the current world record holder, as well the \vg*{plays}[world record] itself
(excluding the level's creator).
Users can express their appreciation of a level
with \vg*{plays}[likes] or \vg*{plays}[boos],
or by leaving \vg*{plays}[comments]: the corresponding variables record
the total numbers of each.
According to \autoref{tab:numeric},
there is a huge spread in these metrics across levels,
and a limited number of most popular levels
generate massive amounts of plays, likes, and comments.
More interesting, 
the median level generates modest, yet non-negligible, engagement
(it was cleared \n{numeric/50p:clears} times,
and played \n{numeric/50p:attempts} times
by \n{numeric/50p:unique_players_and_versus} users,
who left \n{numeric/50p:likes} like,
\n{numeric/50p:boos} boo,
and \n{numeric/50p:num_comments} comments);
and only a small minority of levels never gets played.\footnote{In no small part thanks to community initiatives such as \href{https://en.wikipedia.org/wiki/Team\_0\%25}{Team 0\%}.}

Most relevant for this paper,
two variables characterize a level's difficulty.
The \vg*{difficulty}[clear rate] is simply the ratio $\var{clears}/\var{attempts}$.
In addition, \smm automatically assigns an ordinal \vg*{difficulty}[difficulty]
rating with four levels: \emph{easy}, \emph{normal}, \emph{expert}, and \emph{super expert}.
\autoref{tab:data-overview} shows that
the median level is rated \emph{normal},
and has a clear rate of \n[2]{numeric/50p:clear_rate_p} (i.e., \n{numeric/50p:clear_rate_p}[100]|).
The formula used by Nintendo to assign difficulty ratings is not known;
however, it is likely based on the clear rate (although it's not \emph{exclusively} based on it, since two levels with the same clear rate may get different difficulty ratings) and is dynamic (i.e., it may change over time).
Intuitively, the clear rate is a more precise assessment of
a level's difficulty, given that it is more fine grained.
However, it's not a perfect measure:
the clear rate does not distinguish between popular 
(which have been played by myriad players with all kinds of skills)
and unpopular levels
(whose clear rate may simply reflect the skills of the small, self-selected
group of players who attempted it);
and, ultimately, ``difficulty'' is partly subjective (e.g., depending on a player's familiarity with the mechanisms used by a game).
Nevertheless, \var{clear rate} remains the best proxy for a level's difficulty
in the metadata;
hence, we'll rely on it in our analysis.

\subsection{User Data}
The user data include
\n{N.user.vars} variables;
as for the level data,
we distilled these down to \n{N.all.user.vars}
variables, which we grouped into the \vg{user} category.
Each user has a unique identifier,
and is associated a
country and a \vg*{user}[region] (roughly corresponding to continents)
of activity.

The system stores a few key metrics of user activity as a \emph{player}:
the total number of levels they \vg*{user}[played]
and successfully \vg*{user}[cleared],
as well as the number of \vg*{user}[attempted] plays,
and how many were \vg*{user}[deaths];\footnote{
  Thus, \var{played} and \var{cleared} refer to \emph{levels},
  whereas \var{attempted} and \var{deaths} refer to individual plays
  of any level.
}
Other variables record a user's \vg*{user}[first clears]
(the number of levels that they cleared before any other player)
and world \vg*{user}[records]
(the number of levels that they cleared faster than any other player).
Then, a few variables record the user's activity
in specific game modes:
their high \vg*{user}[score]
in the endless challenge
(where a user plays random levels of a certain difficulty
until they run out of lives)
for each difficulty;
and their \vg*{user}[versus rating] in multiplayer versus
(where four users race to reach the end of a level).

Finally, there are a few key metrics
of a user's activity as a \emph{maker}:
the system assigns \vg*{user}[maker points],
which reward a maker's activity and achievements;
the number of levels a user \vg*{user}[uploaded],\footnote{
  At any given time, a user can share
  \n{numeric/max:uploaded_levels} levels maximum; however,
  one may delete some levels and replace them with new uploads.
  Variable \var{uploaded} records the current number of
  published levels, and hence it is capped at \n{numeric/max:uploaded_levels}.
}
and whether they allow other users to leave comments
(variable \vg*{user}[comments?]);
and the total number of likes the maker's levels
received (variable \vg*{user}[maker likes]).

Just like for level data,
\autoref{tab:numeric} shows that there is a large spread
in user data as well,
and that most users have a non-trivial activity record.
The median user cleared \n{numeric/50p:courses_cleared} levels,
and claimed first clear on \n{numeric/50p:first_clears}
and world record on \n{numeric/50p:world_records}.
Remarkably, an ample majority of users are also \emph{makers}:
the median user collected \n{numeric/50p:maker_points} maker points
by uploading \n{numeric/50p:uploaded_levels} levels.

\section{Regression Analysis of the Data}
\label{sec:regression}

In order to quantitatively analyze which factors affect
a level's difficulty, we fit
on the level and user data
generalized linear regression models of the general form:
\begin{equation}
  \begin{aligned}
    \var{clear rate}_i &\sim\ \dist{Normal}(\mu_i, \sigma) 
    \\
    \log(\mu_i) &=\ \alpha + \sum_{\var{v} \in \vars} \coef{v} \var{v}_i
  \end{aligned}
  \label{eq:regression-model}
\end{equation}
In all models, variable \var{clear rate} is used as outcome,
and other variables in \vars\ are used as predictors; 
the logarithmic link function ensures that the predicted \var{clear rate}
is always a nonnegative value.

Fitting even a simple model such as \eqref{eq:regression-model}
on \smm's massive dataset 
is challenging even with plenty of memory, CPUs, and disk space;
this restricted the techniques and models that we could use:
\begin{enumerate*}
\item classic frequentist fitting algorithms
  instead of more robust, yet computationally demanding
  Bayesian simulation-based techniques~\cite{FTF-TOSEM21-Bayes-guidelines};
\item a normal distribution as likelihood,
  instead of more precise choices 
  (such as a beta distribution, which constrains the outcome to
  be a value over $[0, 1]$, like \var{clear rate})
  that use less efficient fitting algorithms.
\end{enumerate*}

\subsection{Choosing Linear Predictors}

A key choice is which variables to include as predictors in \eqref{eq:regression-model}.
Again for practical performance reasons,
we have to exclude a few level and user variables:
a level's id, upload timestamp, clear conditions, and ids of the user
who first cleared the level and of the one who holds the world record;
and a user's id and country.
All these variables are nominal with a very large number of levels;
since an $\ell$-level nominal variable is modeled
as $\ell - 1$ binary indicator variables,
such variable would effectively blow up the number of variables in \vars,
thus rendering \eqref{eq:regression-model} intractable.

Apart from this restriction, we include as many variables as possible,
since we are only interested in \emph{associations}
and \emph{prediction} (as opposed to causal relations,
which we'll briefly analyze separately later).
This angle also justifies including variables
that correlate closely (i.e., they exhibit multicollinearity),
as long as they do not introduce convergence problems.
Based on these principles,
we consider two models $m_\ell$ and $m_u$.
Model $m_\ell$ 
includes as predictors in $\vars_{\ell}$
all \n{N.fml.all.both.variables} viable level variables:
\var{difficulty},\footnote{
  Interestingly, omitting \var{difficulty} from the model
  incurs convergence problems and leads to a fit with
  poor predictive capabilities.
}
\var{clears}, \var{attempts},
\var{likes}, \var{boos}, \var{players}, \var{world record}, \var{upload time}, \var{comments}, \var{timer}, \var{autoscroll}, \var{style}, \var{theme}, \var{version}, \var{upload attempts}.
Model $m_u$ tries to include, on top of the level data,
the user data about each level's maker.
Extending $m_\ell$ with all viable user variables
incurs convergence problems;
as a workaround, we excluded the three level variables
with the smallest effect in $m_\ell$
(\var{players}, \var{upload time}, \var{timer})
and added all \numprint{16} viable user variables in $\vars[U]$:
\var{played}, \var{cleared}, \var{attempted}, \var{deaths}, \var{maker points},
the high \var{score} in endless for each difficulty,
the \var{versus rating}, \var{first clears}, \var{world records},
\var{uploaded} levels, \var{region}, \var{comments?}, and \var{maker likes}.
In all, $m_u$ uses the \n{N.user.vars} predictors in $\vars_{u} = \vars_{\ell} \setminus \{\var{players}, \var{upload time}, \var{timer}\} \cup \vars[U]$.

\begin{figure*}[!htb]
  \centering
  \includegraphics[width=0.75\linewidth]{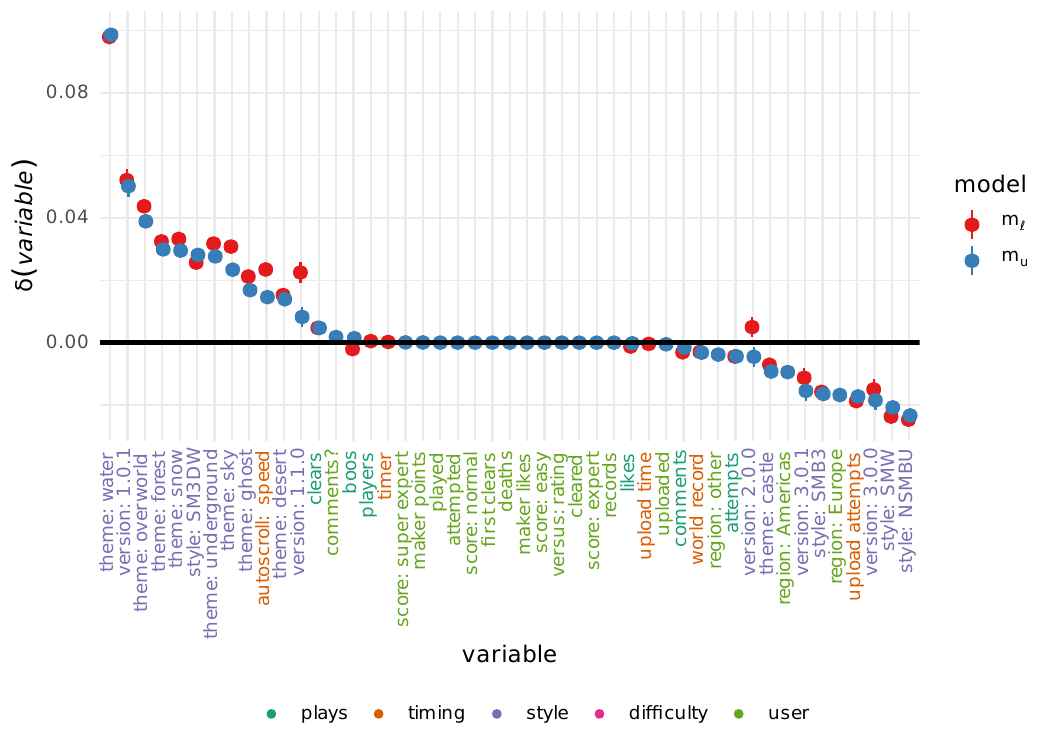}
  \caption{For each regression variable \var{v},
    the value of $\exp(\coef{v})-1$, where \coef{v} is \var{v}'s
    coefficient in \textcolor{mlcol}{model $m_\ell$}
    and in \textcolor{mucol}{model $m_u$}.}
    \label{fig:coefs-mlmu}
\end{figure*}

\begin{figure*}[!htb]
  \centering
  \includegraphics[width=0.7\linewidth]{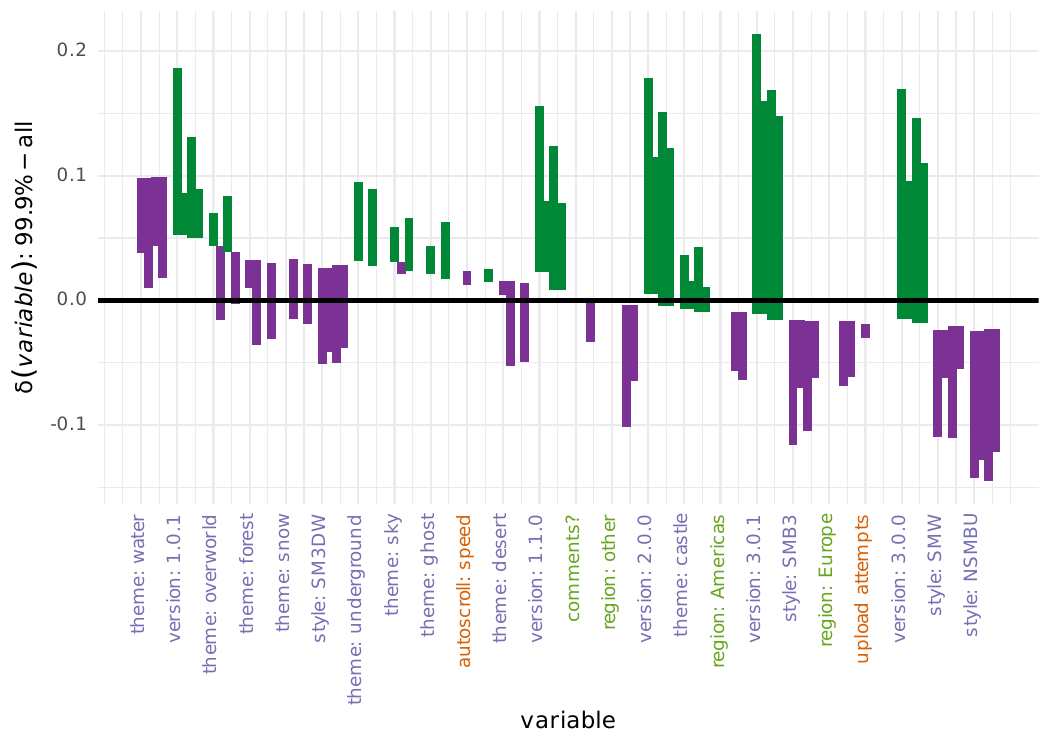}
  \caption{Each bar is the difference $\ddiff{m}{d}[v]$ of $\dcoef[v]$
    fitted on the 99.9\% most popular levels and fitted on all level data.
    Bars \plotbar[positive] represent positive differences and
    bars \plotbar[negative] negative ones.
    For each \var{v}, there are (up to) four bars:
    $\ddiff{\ell}{m}[v]$, $\ddiff{\ell}{p}[v]$, $\ddiff{u}{m}[v]$, $\ddiff{u}{p}[v]$.}
    \label{fig:coefs-diffs}
\end{figure*}

\subsection{Regression Analysis of All Data}
\label{sec:analysis:all}

After fitting models $m_\ell$ and $m_u$ on the \smm data,
we can interpret its coefficients $\beta$ as
strength of association between a predictor variable
and the clear rate.
For a set of $n$ variables $\var{v}^1, \ldots, \var{v}^n$,
a datapoint $\vec{p}$ is an $n$-tuple $\langle p^1, \ldots, p^n \rangle$
of values for each variable.
Consider two datapoints $\vec{p}, \vec{q}$ for the same set of $n$ variables
such that $p^i = q^i$ for all components $i$, except $p^k = q^k + x$.
According to model \eqref{eq:regression-model},
the ratio of the expected value of the clear rate
of a level with data $\vec{p}$ over the expected value of
the clear rate of a level with data $\vec{q}$ is thus:
\begin{equation}
  \small
  \frac{\mu^p}{\mu^q}
  =
  \frac
  {\exp(\alpha + \beta_1 p^1 + \cdots)}
  {\exp(\alpha + \beta_1 q^1 + \cdots)}
  =
  \exp(\beta_k (p^k - q^k))
  =
  \exp^x(\beta_k)
  \label{eq:ratio}
\end{equation}
This leads to the following interpretation
for numeric and for nominal variables:
\begin{description}
\item[numeric:]  the expected ratio of change of \var{clear rate}
  associated with a \emph{unit} change of a numeric variable \var{v}
  is $\exp(\coef{v})$.
\item[nominal:] since a nominal variable \var{v}
  with $n$ levels $x_1, \ldots, x_n$
  is modeled as $n - 1$ binary variables $\var{v}_{2}, \ldots, \var{v}_{n}$,
  the ratio of change of \var{clear rate}
  associated with a change of \var{v} from the baseline\footnote{
    The baseline levels of nominal variables in our models are:
    \emph{easy} for \var{difficulty},
    \emph{airship} for \var{theme},
    \emph{1.0.0} for \var{version},
    \emph{Asia} for \var{region}.
    }
  level $x_1$ to level $x_{k > 1}$ is $\exp(\beta_{\var{v}_{k}})$.
\end{description}
In the following, we will actually analyze, for each variable $\var{v}$,
the value $\dcoef[v] = \exp(\coef{v}) - 1$.
Since $\exp(\coef{v})$ is a ratio of means,
\dcoef[v] can be readily interpreted as the fractional increase
(if positive) or decrease (if negative) of \var{clear rate} associated
with a unit change in that variable.

Let's start with variable \vg*{difficulty}[difficulty]:
in model $m_\ell$,
$\dcoef[normal] = \n[2]{coefs.both.user/difficulty_nameNormal/fml.all.both}$,
$\dcoef[expert] = \n[2]{coefs.both.user/difficulty_nameExpert/fml.all.both}$,
$\dcoef[super expert] = \n[2]{coefs.both.user/difficulty_nameSuper expert/fml.all.both}$.
This means that, on average, a \emph{normal} level has a clear rate that is \n[0]{coefs.both.user/difficulty_nameNormal/fml.all.both}[-100]|
lower than an \emph{easy} level;
an \emph{expert} level's clear rate is 
\n[0]{coefs.both.user/difficulty_nameExpert/fml.all.both}[-100]| lower;
and a \emph{super expert} level's clear rate is
\n[0]{coefs.both.user/difficulty_nameSuper expert/fml.all.both}[-100]| lower.
This is reasonable, as it is unusual to find a
super expert level with a clear rate that is much higher than 2\%.

Given that \vg*{difficulty}[difficulty] is based on the clear rate,
it is unsurprising that the corresponding \dcoef{}s have a disproportionate
association with the outcome.
The rest of the analysis, in particular \autoref{fig:coefs-mlmu},
focuses on the \dcoef{}s for all \emph{other} predictors in models
$m_\ell$ and $m_u$.\footnote{
  This does not mean that we remove \vg*{difficulty}[difficulty] from our models,
  but simply that we zoom in on the other predictors.
  }

  Most \vg*{user} variables have a negligible association with the outcome;
  in fact, model $m_u$ has worse (higher) AIC score~\cite{stat-rethinking} than $m_\ell$ despite
  including many more variables,
  which suggests that $m_\ell$ outperforms $m_u$ in predictive power.
The only \vg*{user} variables with a noticeable association
  ($|\dcoef| > 10^{-3}$) 
  with the outcome are \vg*{user}[region] 
  and \vg*{user}[comments?].
  Thus, makers based in Asia tend to build easier levels than makers in other regions;
and makers who allow others to leave comments on their levels
  tend to design easier levels.

Among variables related to \vg{timing},
  the \vg*{timing}[timer] has negligible impact---probably because
  most levels stick to the default timer.
  The \vg*{timing}[autoscroll speed] is instead associated with higher clear rate;
  this is probably just a result of most levels not using autoscroll,
  and hence does not provide any clear insight.
  In contrast, the \vg*{timing}[upload time], \vg*{timing}[world record], and \vg*{timing}[upload attempts] are all associated with lower clear rates:
  naturally, if a level took a long time to clear it is usually harder.
  Variables \vg*{plays}[comments] and \vg*{plays}[attempts]
  are also all associated with lower clear rates,
  which would seem to indicate that levels that generate a lot of plays
  tend to be harder.
  Then, variables \vg*{plays}[likes] and \vg*{plays}[boos] do not have a consistent
  association one way or another;
  whereas variable \vg*{plays}[clears] is associated with easier levels,
  as these would easily pile up a lot of clears.

  The impact of variables of group \vg{style} is mixed.
  \autoref{fig:coefs-mlmu} suggests the following ranking
  of game \vg*{style}[style],
  from easier to harder: SM3DW, SMB1, SMB3, SMW, NSMBU.
  Each style from SMB1 to NSMBU extends the game with new mechanics
  (e.g., holding objects, spin jumps, wall jumps),
  which provide additional ways of building challenging levels.
  However, SM3DW also offers its own distinct game mechanics,
  which is hard to square with it being the style
  with the easier levels on average.\footnote{On the other hand,
    trivial ``refreshing'' levels abound in SM3DW, which might account for part of the difference.}
  Newer \vg*{style}[version]s of \smm
  are associated with harder levels;
  in this case too, this may reflect the additional items and features
  that have been added with every major version,
  as well as a more mature, skilled community of makers.
  Most \vg*{style}[theme]s are associated with easier levels
  than those with the baseline \emph{airship} theme;
  the exception is the \emph{castle} theme,
  which is the style associated with the hardest levels.
  We might speculate that the features available in these themes
  (e.g., a \emph{castle}'s lava floor or an \emph{airship}'s bobbing camera)
  may support introducing additional challenges.
  Somewhat surprisingly, theme \emph{water} has the strongest
  association with easier levels---despite the perception that water levels can be quite challenging
  due to the mechanics of swimming.

\ifextended
A limitation of the interpretation of the \dcoef{}s
is that \dcoef[v] characterizes the change associated with
a \emph{unit} change in variable \var{v},
but different numeric variables may range over quite different scales
(as we have seen in \autoref{tab:numeric}).
To account for this fact, we introduce the metric
$\Dcoef[v] = \exp^{\mad[v]}(\coef{v})$, where $\mad[v]$
is the mean absolute difference of variable \var{v}
in the \smm dataset.\footnote{Since the dataset is too large to compute
  all pairwise differences, we consider a random sample of the 0.1\%
  of all datapoints (i.e., \n[0]{N.levels}[0.001] levels).
}
According to \eqref{eq:ratio}, \Dcoef[v] is thus the expected ratio
of change of \var{clear rate} associated with a mean change of
\var{v}.
While for most variables \dcoef[v] and \Dcoef[v] present a consistent picture,
\Dcoef[v] suggests a stronger association in practice for
variables \vg*{timing}[timer], \vg*{timing}[upload time],
\vg*{user}[maker points], \vg*{user}[played], \vg*{user}[attempted].
The $\dcoef$s of all these variables are practically insignificant,
whereas their $\Dcoef$s suggest that, on average,
\vg*{timing}[upload time] accounts for a
\n[0]{average.effects/upload_time/fml.all.both:mean_delta}[100]|
difference in \var{clear rate};
\vg*{user}[played] and \vg*{user}[attempted]
for \n[0]{average.effects/courses_played/fml.user.all:mean_delta}[100]|;
\vg*{timing}[timer] for
\n[0]{average.effects/timer/fml.all.both:mean_delta}[100]|;
and
\vg*{user}[maker points] for 
\n[0]{average.effects/maker_points/fml.user.all:mean_delta}[100]|.
Thus,
levels with a much longer timer tend to have lower clear rates
(possibly because levels requiring long to clear
put off more players);
and
makers who are particularly active in
earning maker points
tend to design easier levels.
The latter observation may indicate that makers with more
experience produce levels with a \emph{better design} and fairness/usability,
which increases the chances that others are motivated and capable
of clearing them.
\fi

\subsection{Regression Analysis of Popular Levels}
\label{sec:analysis:top}

\autoref{fig:coefs-mlmu}
shows that the predictions of models $m_\ell$ and $m_u$
are largely consistent (except possibly for certain \vg*{style}[version]s).
Does the relative impact of each variable on the clear rate
change if we focus on a smaller set of \emph{popular} levels?
To answer this question, we introduce two datasets:
$D^p$ includes all \n{N.99.9.players} levels whose number of \vg*{plays}[players]
is above the 99.9\% percentile;
and $D^m$ all \n{N.99.9.makers} levels whose maker's \vg*{user}[maker points]
are above the 99.9\% percentile.
For each variable \var{v}, \ddiff{m}{d}[v] denotes the difference
between \dcoef[v] in model $m_m$ fitted on $D^d$
and the same model fitted on all data.

For several variables \var{v}, \ddiff{m}{d}[v] is
smaller than 0.01, and hence negligible.
\autoref{fig:coefs-diffs} plots
the non-negligible values of \ddiff{m}{d}[v].
For several \vg*{style}[theme]s and for the \vg*{style}[SM3DW] style
the difference crosses
the zero line, which means that we see opposite trends in popular
vs.\ all levels.
In particular, popular SM3DW levels tend to be harder than popular SMB1
levels, which corroborates the conjecture that the variety of mechanics
available in SM3DW support experienced makers to build harder levels.
A similar observation holds for certain themes (\emph{water}, \emph{forest}, \emph{snow}, \emph{underground}, and \emph{desert}) that tend to be harder to beat
in popular levels than in all levels.
Conversely, popular levels using theme \emph{castle} tend to be easier;
and the changes are inconsistent for the other themes.
For most variables, however, the difference
\ddiff{m}{d}[v]
does not cross the zero line; 
the most popular levels often present
the same kind of associations
between \var{v} and the level's clear rate,
but in stronger form.
For example, the difficulty ranking of styles \emph{SMB3}$<$\emph{SMW}$<$\emph{NSMBU}
still roughly holds for popular levels, but each
has an overall stronger association with the clear rate.

For clarity,
\autoref{fig:coefs-diffs} omits variables related to \vg*{difficulty}[difficulty],
which range over quite different values than all other variables.
While the strong association between difficulty rating and clear rate remains
for \vg*{difficulty}[difficulty] variables,
\ddiff{\ell}{m} and \ddiff{u}{m} are usually negative,
whereas \ddiff{\ell}{p} and \ddiff{u}{p} are positive.
Thus, within the same difficulty rating,
popular makers tend to produce harder levels,
but the most played levels tend to be a bit easier.

\ifextended
\subsection{Causal Regression Models}
\label{sec:causal}

All the analyses presented so far
can only report \emph{associations} among variables, since they target \emph{observational}
data---and not controlled experiments, which can tease out
genuine \emph{causal} relations.
In this section, we outline\footnote{
  For lack of space, details are only presented in the
  replication package.
}
the results of an analysis of the same observational data
designed in a way that it accounts for (some of) the
actual causal relations among variables.
Following Pearl's techniques~\cite{pearl09a@reason,FTF-TOSEM23-causality-CodeJam},
we built a structural model of the possible cause/effect relations among
some of the \smm data variables.
The model suggests a so-called \emph{adjustment set}:
a set of variables $\vars[C]$ that we can use as $\vars$ in \eqref{eq:regression-model}.
The resulting regression model $m_c$ corrects for possible non-causal biases in the
estimates of the \dcoef[v].
According to our hypotheses, model $m_c$ only includes predictors
\vg*{style}[style], \vg*{style}[theme], \vg*{style}[version],
and \vg*{timing}[upload attempts].
Thus, while $m_c$ likely has less non-causal \emph{bias}
than $m_\ell$ and $m_u$, it is also much less \emph{precise}
in predicting the outcome.
Therefore, we should not be surprised that the coefficient
estimates can differ conspicuously,
as the precise results of analyzing $m_c$
are hardly directly comparable to those of $m_\ell$ and $m_u$.

Despite these limitations,
we can still try to compare the \emph{sign} of \dcoef[v] in $m_c$
against that in the other models.
In a nutshell, the sign agrees for most shared variables
with two exceptions:
\begin{enumerate*}
\item The \vg*{style}[theme] \emph{ghost house} is associated with
  a positive \dcoef\ in $m_\ell$/$m_u$ but with a negative \dcoef\ in $m_c$.
  Thus, the association with easier levels may be a result of confounding
  and not of actual causal links.

\item The \dcoef{}s of most \vg*{style}[version] variables flip sign
  (either way) in $m_c$ compared to $m_\ell$/$m_u$.
\end{enumerate*}
Given that variable \vg*{style}[version]
often had an inconsistent association (e.g., 2.0.0 in $m_\ell$ vs.~$m_u$),
the only tenable conclusion is
that the \smm version has at most an indirect, weak
relation with a level's clear rate,
which is easily confounded.
\fi

\section{NLP Analysis}
\label{sec:nlp}

We leverage NLP (natural language processing) techniques to analyze how level titles and descriptions 
(\autoref{sec:nlp:titledesc}), and comments (\autoref{sec:nlp:comments}) relate to a level's \vg*{difficulty}[difficulty].

\subsection{Titles and Descriptions}
\label{sec:nlp:titledesc}
 
The title and the description of a level represent the creator's
highlights of the level's content. To study how 
this correlates with the level's \vg*{difficulty}[difficulty],
we leverage \bertopic, a 
topic modeling approach~\cite{grootendorst2022bertopic}. \bertopic
extracts latent topics from a collection of documents; in our case, each level is a document
consisting of the concatenation of the level's title and description. The base
\bertopic algorithm uses pre-trained transformer-based language models to build
 document embeddings, which are clustered by similarity to derive latent topics. Then, \bertopic
 derives topic representations according to their class-based term frequency-inverse document frequency (TF-IDF). We configured \bertopic to use words and bigrams as topics;
 in practice, a topic is a collection of words that
 tend to occur together in several documents.
 In the following, ``topic-word'' refers to any of the words that
 characterize a topic.

Consistently with the game's worldwide popularity,
\smm titles and descriptions are written in many different languages.
We focus on English text, which is the most widely used language.
Since the \smm dataset has no information about language,
we use the \textsc{Lingua} language detector~\footnote{See \url{https://github.com/pemistahl/lingua-py}.}, obtaining a total of \n{sent:titledesc:N.levels} levels with title and descriptions in English.

To understand which topics characterize each difficulty, we 
calculate the relative frequency of a topic $t$ for each \vg*{difficulty}[difficulty] $d$ (easy, normal, expert, super expert)
as the 
fraction of levels with difficulty $d$
whose title/description matches topic $t$.
Based on this metric,
we call topic $t$
a \emph{characterizing topics} for \vg*{difficulty}[difficulty] $d$
if $t$'s frequency in levels of difficulty $d$
is higher than its frequency in levels of other difficulties.
Of the total 179 topics extracted 
with \bertopic, 52 are characterizing for \emph{super expert}, 20 for \emph{expert}, 37 for \emph{normal}, and 70 for \emph{easy}.
\autoref{tab:desc:topics} shows the top-5 characterizing topics for the \emph{super expert} and \emph{easy} difficulties. 

\begin{table}[h]
  \centering
  \scriptsize
    \setlength{\tabcolsep}{4.5pt}
\begin{tabular}{llccccccccccccccccccccccc}
  \toprule
  \multicolumn{2}{c}{topic} &
  \multicolumn{4}{c}{frequency per \textcolor[HTML]{E7298A}{difficulty}} \\
  \cmidrule(l{3pt}r{3pt}){1-2} \cmidrule(l{3pt}r{3pt}){3-6} 
  \# & \multicolumn{1}{c}{top-4 topic words} & \rotatebox{-90}{easy} & \rotatebox{-90}{normal} & \rotatebox{-90}{expert} & \rotatebox{-90}{super expert} \\
  \toprule
  \#15 & [speed, speedrun, run, seconds]  & \n[2]{topics:desc:top_n_superexpert:15:Easy_p}[100]| & \n[2]{topics:desc:top_n_superexpert:15:Normal_p}[100]| & \n[2]{topics:desc:top_n_superexpert:15:Expert_p}[100]| & \n[2]{topics:desc:top_n_superexpert:15:Super_Expert_p}[100]| \\ 
  \#13 & [jump, jumps, jumping, long]  & \n[2]{topics:desc:top_n_superexpert:13:Easy_p}[100]| & \n[2]{topics:desc:top_n_superexpert:13:Normal_p}[100]| & \n[2]{topics:desc:top_n_superexpert:13:Expert_p}[100]| & \n[2]{topics:desc:top_n_superexpert:13:Super_Expert_p}[100]| \\ 
  \#32 & [level, lol, sorry, little]  & \n[2]{topics:desc:top_n_superexpert:32:Easy_p}[100]| & \n[2]{topics:desc:top_n_superexpert:32:Normal_p}[100]| & \n[2]{topics:desc:top_n_superexpert:32:Expert_p}[100]| & \n[2]{topics:desc:top_n_superexpert:32:Super_Expert_p}[100]| \\ 
  \#56 & [practice, tricks, jumps, basic]  & \n[2]{topics:desc:top_n_superexpert:56:Easy_p}[100]| & \n[2]{topics:desc:top_n_superexpert:56:Normal_p}[100]| & \n[2]{topics:desc:top_n_superexpert:56:Expert_p}[100]| & \n[2]{topics:desc:top_n_superexpert:56:Super_Expert_p}[100]| \\ 
  \#25 & [ride, spin, run, victory]  & \n[2]{topics:desc:top_n_superexpert:25:Easy_p}[100]| & \n[2]{topics:desc:top_n_superexpert:25:Normal_p}[100]| & \n[2]{topics:desc:top_n_superexpert:25:Expert_p}[100]| & \n[2]{topics:desc:top_n_superexpert:25:Super_Expert_p}[100]| \\
  \bottomrule
  \#0 & [maker, super, bros, 11]  & \n[2]{topics:desc:top_n_easy:0:Easy_p}[100]| & \n[2]{topics:desc:top_n_easy:0:Normal_p}[100]| & \n[2]{topics:desc:top_n_easy:0:Expert_p}[100]| & \n[2]{topics:desc:top_n_easy:0:Super_Expert_p}[100]| \\ 
  \#5 & [level, level easy, easy level]  & \n[2]{topics:desc:top_n_easy:5:Easy_p}[100]| & \n[2]{topics:desc:top_n_easy:5:Normal_p}[100]| & \n[2]{topics:desc:top_n_easy:5:Expert_p}[100]| & \n[2]{topics:desc:top_n_easy:5:Super_Expert_p}[100]| \\ 
  \#30 & [world, 11, 12, 13]  & \n[2]{topics:desc:top_n_easy:30:Easy_p}[100]| & \n[2]{topics:desc:top_n_easy:30:Normal_p}[100]| & \n[2]{topics:desc:top_n_easy:30:Expert_p}[100]| & \n[2]{topics:desc:top_n_easy:30:Super_Expert_p}[100]| \\ 
  \#16 & [hard, easy, try, impossible]  & \n[2]{topics:desc:top_n_easy:16:Easy_p}[100]| & \n[2]{topics:desc:top_n_easy:16:Normal_p}[100]| & \n[2]{topics:desc:top_n_easy:16:Expert_p}[100]| & \n[2]{topics:desc:top_n_easy:16:Super_Expert_p}[100]| \\ 
  \#40 & [water, dangerous, madness, life]  & \n[2]{topics:desc:top_n_easy:40:Easy_p}[100]| & \n[2]{topics:desc:top_n_easy:40:Normal_p}[100]| & \n[2]{topics:desc:top_n_easy:40:Expert_p}[100]| & \n[2]{topics:desc:top_n_easy:40:Super_Expert_p}[100]| \\
  \bottomrule
\end{tabular}
\setlength{\tabcolsep}{6pt}
   \caption{Top-5 characterizing topics, by frequency, in titles and descriptions of \emph{super expert} (top) and \emph{easy} (bottom) levels.}
  \label{tab:desc:topics}
\end{table} 

\titledsubpar{Topics characterizing super expert levels.} The most frequent topic-word characterizing \emph{super expert} levels is
\textbf{speed}. Overall, \textbf{speed} is a topic-word in $4$ topics,
of which $3$ are characterizing for \emph{super expert}.
Similarly, \textbf{jump} is a topic-word in $14$ topics,
of which $13$ are characterizing for \emph{super expert}. All 
the $4$ topics containing \textbf{spin} as topic-word,
and all the $5$ topics containing \textbf{practice}, 
are also characterizing for \emph{super expert}. These probably indicate 
levels that let players practice and master a variety of \smm's challenging game
mechanics; in fact, these topics also include topic-words such as \textbf{jump}, \textbf{spin}, \textbf{fly}, and \textbf{tricks}.

\titledsubpar{Topics characterizing easy levels.} The most frequent topic-words characterizing 
\emph{easy} levels are very generic terms like \textbf{maker} and
\textbf{world}, which probably just means that easy levels tend to have generic titles and descriptions. 
Also self-explanatory is the second-most occurring topic-word: \textbf{easy};
among the $14$ topics that contains it, $7$ are characterizing for the \emph{easy}
\vg*{difficulty}[difficulty].
One seemingly unexpected topic-word is the number \textbf{11};
this is actually due to the way \bertopic encodes topic-words
by removing non-alphanumeric characters.
Thus, \textbf{11} actually stands for \textbf{1--1}, which is
the canonical way of referring to the first level of a multi-level world and, in particular, to the iconic \emph{World 1--1} in the original Super
Mario Bros.
Replicating such classic level in \smm, often with unique twists,
is its own sub-genre;\footnote{See for example \url{https://www.youtube.com/watch?v=PhyG0s9tJaM}}
our findings indicate that this is particularly popular with easy level---although 1--1 variants are found at all difficulties.
Other topic-words consisting of two-digit numbers (e.g., 12, 13)
have a similar explanation:
they stand for \emph{world}--\emph{level} identifiers (e.g., 1--2, 1--3).
These are either replicas of original Super Mario Bros.\ levels,
or identify levels that belong to \emph{Super Worlds}.\footnote{See \url{https://supermariomaker2.fandom.com/wiki/Super_Worlds}}
Some of the remaining topics characterizing easy levels
seem to contradict the levels' difficulty ranking,
as they include topic-words such as \textbf{hard}, \textbf{impossible},
\textbf{dangerous}, and \textbf{madness}.
A possible explanation is the makers' taste for deliberately misleading
``trolling''\footnote{\url{https://docs.google.com/document/d/13ZoqeblLs45HuEfTtsOrq6X0LAuEnA8nB721_doxE38}} titles, whereupon trivial levels are marked ``100\% impossible'' and very hard ones are titled ``easy''.

\subsection{Comments}
\label{sec:nlp:comments}

While titles and descriptions express the maker's point of view on their levels,
comments allow any players 
to voice their opinion. 
We first look at the \emph{sentiment} of level comments,
and then at their \emph{topics}.
As in \autoref{sec:nlp:titledesc},
we only consider comments written in English, as detected by \textsc{Lingua}.
To have a meaningful set of comments for each level,
we only consider levels with at least 6 comments (3\% of all levels), corresponding to a total of \n{sent:comments:N.levels} levels and 
\n{sent:comments:N.comments} comments.
Let $C(\ell)$ denote the set of comments of a level~$\ell$.

\titledpar{Sentiment.} We analyze the sentiments of the comments
by using Barbieri et al.'s~\cite{barbieri-etal-2020-tweeteval} transformer-based pipeline. For a comment $c$, the model estimates its sentiment $S(c)$
as a triple $\langle S_-(c), S_0(c), S_+(c)  \rangle$ of scores representing
the fraction of \emph{negative} $-$, \emph{neutral} $0$, and \emph{positive} $+$
sentiment, where $S_i(c) \in [0, 1]$ and $\sum_{i} S_i(c) = 1$,
$i \in \{-, 0, +\} \equiv K$.
Consider the following derived metrics,
summarized in \autoref{tab:comments:sent}:
\begin{itemize}
\item The \emph{average sentiment} $S_i(\ell)$ of $\ell$
  is the mean of $S_i(c)$ over all comments $c$ of level $\ell$.
\item The \emph{dominant sentiment} $D(c) \in K$
  is the sentiment with the highest score in $S(c)$:
  $D(c) = \argmax_{{i}} S_i(c)$.
\item $C_i(\ell) \subseteq C(\ell)$ is the set of all comments
  of $\ell$ whose dominant sentiment is $i \in K$. 
\item $\SR_i(\ell)$ is the fraction
  $|C_i(\ell)| / |C(\ell)|$
  of $\ell$'s comments whose
  dominant sentiment is $i \in K$.
\item Similarly, $\overline{S_i(X)}$ and $\overline{\SR_i(X)}$
  denote the mean of $S_i(\ell)$ and $\SR_i(\ell)$ for all levels
  of difficulty in the set $X$ of difficulties.
\end{itemize}

\begin{table}[!htb]
  \centering
  \scriptsize
    \setlength{\tabcolsep}{4pt}
\begin{tabular}{c *{4}{rr}}
  \toprule
  \multicolumn{1}{c}{\textcolor[HTML]{E7298A}{difficulty}} & \multicolumn{2}{c}{$d[S_i^x]$} & \multicolumn{2}{c}{$d[\SR_i^x]$} & \multicolumn{2}{c}{$\overline{S_i(\{x\})}$}  & \multicolumn{2}{c}{$\overline{\SR_i(\{x\})}$}\\

 $x$ & $-$ & $+$ & $-$ & $+$  & $-$ & $+$ & $-$ & $+$\\
  \cmidrule(l{3pt}r{3pt}){1-1} \cmidrule(l{3pt}r{3pt}){2-3}  \cmidrule(l{3pt}r{3pt}){4-5} \cmidrule(l{3pt}r{3pt}){6-7} \cmidrule(l{3pt}r{3pt}){8-9} 
  easy & \n[2]{d_S_i_comment:easy:neg} & \n[2]{d_S_i_comment:easy:pos} & \n[2]{d_SR_i_comment:easy:neg} & \n[2]{d_SR_i_comment:easy:pos} & \n[2]{avg_S_i_level:easy:neg} & \n[2]{avg_S_i_level:easy:pos} & \n[2]{avg_SR_i:easy:neg} & \n[2]{avg_SR_i:easy:pos}\\
  normal & \n[2]{d_S_i_comment:normal:neg} & \n[2]{d_S_i_comment:normal:pos} & \n[2]{d_SR_i_comment:normal:neg} & \n[2]{d_SR_i_comment:normal:pos} & \n[2]{avg_S_i_level:normal:neg} & \n[2]{avg_S_i_level:normal:pos} & \n[2]{avg_SR_i:normal:neg} & \n[2]{avg_SR_i:normal:pos} \\
  expert & \n[2]{d_S_i_comment:expert:neg} & \n[2]{d_S_i_comment:expert:pos} & \n[2]{d_SR_i_comment:expert:neg} & \n[2]{d_SR_i_comment:expert:pos} & \n[2]{avg_S_i_level:expert:neg} & \n[2]{avg_S_i_level:expert:pos} & \n[2]{avg_SR_i:expert:neg} & \n[2]{avg_SR_i:expert:pos}\\
  super expert & \n[2]{d_S_i_comment:superexpert:neg} & \n[2]{d_S_i_comment:superexpert:pos} & \n[2]{d_SR_i_comment:superexpert:neg} & \n[2]{d_SR_i_comment:superexpert:pos} & \n[2]{avg_S_i_level:superexpert:neg} & \n[2]{avg_S_i_level:superexpert:pos}& \n[2]{avg_SR_i:superexpert:neg} & \n[2]{avg_SR_i:superexpert:pos} \\
  \bottomrule
  \end{tabular}
\setlength{\tabcolsep}{6pt}
     \caption{An overview of the sentiments in level comments.
      The table shows the Cliff's delta of the relationship between a difficulty and all other difficulties,
      as well as the mean of the average sentiment $S_i(\ell)$
      and of the fraction of dominant sentiment $\SR_i(\ell)$
      over all levels $\ell$ of each difficulty $x$.}
  \label{tab:comments:sent}
\end{table}

A trend visible in \autoref{tab:comments:sent}
is that \emph{super expert} level comments
display a higher negative average sentiment than other difficulties;
this holds both for the $S_{-}$ and the $\SR_{-}$ metrics,
that is whether we aggregate by average or by dominant sentiment.
To quantify this observation,
we compute Cliff's delta---a non-parametric effect size,
suitable to quantify how often the values
in one set are larger than the values in another, independent set,
without assumptions about their underlying distributions.
Let $d[S_i^x]$ denote Cliff's delta of the differences
between $\overline{S_i(\{x\})}$ and $\overline{S_i(\{y \neq x\})}$, i.e., between the average $S_i$ for all levels of difficulty $x$ and the average $S_i$ for all other levels;
similarly, $d[\SR_i^x]$ denotes Cliff's delta of the corresponding difference
of metric $\SR_i$.
\autoref{tab:comments:sent} shows that
$d[S_-^{\text{super expert}}] = \n[2]{d_S_i_comment:superexpert:neg}$
and $d[\SR_-^{\text{super expert}}] = \n[2]{d_SR_i_comment:superexpert:neg}$,
Such effect sizes are usually considered \emph{small} but not negligible~\cite{Romano:2006};\footnote{Null-hypothesis statistical testing is uninformative in this case (possibly in general~\cite{pvalue-cohen}),
  since the sheer amount of data leads to minuscule $p$-values.
}
in contrast, $d[S_i^x]$ and $d[\SR_i^x]$ are \emph{negligible}
for all other difficulties $x$ other than \emph{super expert},
and in all ratings corresponding to positive sentiment, 
which supports our observation that players of super expert
 levels tend to express more negative sentiments.

\titledpar{Topics.}
We analyzed English user comments by applying \bertopic as in \autoref{sec:nlp:titledesc},
aggregating comment topic occurrences by \emph{level};
thus, a topic $t$ has occurrence frequency $\tau\%$ for a certain difficulty $x$
if $\tau\%$ of the $x$-difficulty levels include at least one comment
with topic $t$.
\autoref{tab:comments:topics}
lists the top-5 characterizing topics for the
\emph{super expert} and \emph{easy} difficulties.

\begin{table}[h]
  \centering
  \scriptsize
    \setlength{\tabcolsep}{4pt}
\begin{tabular}{llrrrrc}
  \toprule
  \multicolumn{2}{c}{topic} &
  \multicolumn{4}{c}{frequency per \textcolor[HTML]{E7298A}{difficulty}} \\
  \cmidrule(l{3pt}r{3pt}){1-2} \cmidrule(l{3pt}r{3pt}){3-6} 
  \# & \multicolumn{1}{c}{top-4 topic words} & \multicolumn{1}{c}{\rotatebox{-90}{easy}} & \multicolumn{1}{c}{\rotatebox{-90}{normal}} & \multicolumn{1}{c}{\rotatebox{-90}{expert}} & \multicolumn{1}{c}{\rotatebox{-90}{super expert}} \\
  \toprule
  \#0 & [level, great level, good level]  & \n[2]{topics:comm:top_n_superexpert:0:Easy_p}[100]| & \n[2]{topics:comm:top_n_superexpert:0:Normal_p}[100]| & \n[2]{topics:comm:top_n_superexpert:0:Expert_p}[100]| & \n[2]{topics:comm:top_n_superexpert:0:Super_Expert_p}[100]| \\ 
  \#5 & [hard, easy, beat, challenge]  & \n[2]{topics:comm:top_n_superexpert:5:Easy_p}[100]| & \n[2]{topics:comm:top_n_superexpert:5:Normal_p}[100]| & \n[2]{topics:comm:top_n_superexpert:5:Expert_p}[100]| & \n[2]{topics:comm:top_n_superexpert:5:Super_Expert_p}[100]| \\ 
  \#10 & [jump, spin, jumping, jumped]  & \n[2]{topics:comm:top_n_superexpert:10:Easy_p}[100]| & \n[2]{topics:comm:top_n_superexpert:10:Normal_p}[100]| & \n[2]{topics:comm:top_n_superexpert:10:Expert_p}[100]| & \n[2]{topics:comm:top_n_superexpert:10:Super_Expert_p}[100]| \\ 
  \#50 & [shell, jump, throw, double]  & \n[2]{topics:comm:top_n_superexpert:50:Easy_p}[100]| & \n[2]{topics:comm:top_n_superexpert:50:Normal_p}[100]| & \n[2]{topics:comm:top_n_superexpert:50:Expert_p}[100]| & \n[2]{topics:comm:top_n_superexpert:50:Super_Expert_p}[100]| \\ 
  \#38 & [troll, bully, mad, lad]  & \n[2]{topics:comm:top_n_superexpert:38:Easy_p}[100]| & \n[2]{topics:comm:top_n_superexpert:38:Normal_p}[100]| & \n[2]{topics:comm:top_n_superexpert:38:Expert_p}[100]| & \n[2]{topics:comm:top_n_superexpert:38:Super_Expert_p}[100]| \\
  \bottomrule
  \#1 & [mario, toad, bros, maker]  & \n[2]{topics:comm:top_n_easy:1:Easy_p}[100]| & \n[2]{topics:comm:top_n_easy:1:Normal_p}[100]| & \n[2]{topics:comm:top_n_easy:1:Expert_p}[100]| & \n[2]{topics:comm:top_n_easy:1:Super_Expert_p}[100]| \\ 
  \#2 & [nintendo, switch, minecraft]  & \n[2]{topics:comm:top_n_easy:2:Easy_p}[100]| & \n[2]{topics:comm:top_n_easy:2:Normal_p}[100]| & \n[2]{topics:comm:top_n_easy:2:Expert_p}[100]| & \n[2]{topics:comm:top_n_easy:2:Super_Expert_p}[100]| \\ 
  \#3 & [course, great course, fun course]  & \n[2]{topics:comm:top_n_easy:3:Easy_p}[100]| & \n[2]{topics:comm:top_n_easy:3:Normal_p}[100]| & \n[2]{topics:comm:top_n_easy:3:Expert_p}[100]| & \n[2]{topics:comm:top_n_easy:3:Super_Expert_p}[100]| \\ 
  \#4 & [awesome, good job, thanks]  & \n[2]{topics:comm:top_n_easy:4:Easy_p}[100]| & \n[2]{topics:comm:top_n_easy:4:Normal_p}[100]| & \n[2]{topics:comm:top_n_easy:4:Expert_p}[100]| & \n[2]{topics:comm:top_n_easy:4:Super_Expert_p}[100]| \\ 
  \#8 & [brain, eyes, pain, dad]  & \n[2]{topics:comm:top_n_easy:8:Easy_p}[100]| & \n[2]{topics:comm:top_n_easy:8:Normal_p}[100]| & \n[2]{topics:comm:top_n_easy:8:Expert_p}[100]| & \n[2]{topics:comm:top_n_easy:8:Super_Expert_p}[100]| \\ 
  \bottomrule
\end{tabular}
\setlength{\tabcolsep}{6pt}
   \caption{Top-5 characterizing topics, by frequency, in comments of \emph{super expert} (top) and \emph{easy} (bottom) levels.}
  \label{tab:comments:topics}
\end{table} 

\titledsubpar{Comment topics characterizing super expert levels.}
The most frequently occurring characterizing topic in super expert levels
seems to denote appreciation (e.g., \textbf{great}/\textbf{good} level);
indeed, the average $+$ sentiment for comments
with this topic is \n[2]{topics:comm:0:sent_pos:avg}, which is comparatively high.
In contrast,
topic 38 indicates dislike,
and the average $-$ sentiment for comments with this topic
is \n[2]{topics:comm:38:sent_neg:avg},
which is also above average.
This finding complements the previous observation about negative sentiments
in super expert comments:
while users are frustrated by trolly or gratuitously hard levels,
they arguably appreciate when the difficulty is fair
and determines a challenging but satisfying level.
The other characterizing topics of super expert level comments
are in line with \autoref{sec:nlp:titledesc}'s analysis,
showing frequent mention of topic-words that suggest
challenging game mechanic---such as \textbf{shell} \textbf{jumps}
and \textbf{double}/triple jumps.

\titledsubpar{Comment topics characterizing easy levels.}
Topics \#3 and \#4 in \autoref{tab:comments:topics}
seem to indicate generic user appreciation of a level;
in fact the average $+$ sentiment for comments with these topics
are \n[2]{topics:comm:3:sent_pos:avg} and \n[2]{topics:comm:4:sent_pos:avg}
respectively---both clearly above average.
In contrast,
topic \#8 leans towards negative;
its average negative sentiment is
\n[2]{topics:comm:8:sent_neg:avg},
clearly above the average negative sentiment of \emph{easy} levels.
Topic-words such as \textbf{brain}, \textbf{eyes}, and \textbf{pain}
suggest that the most common criticism of easy levels
targets those that make an unnecessary, excessive usage
of flashy
effects,\footnote{\url{https://supermariomaker2.fandom.com/wiki/Sound_Effects}}
which can significantly deteriorate the user experience.
Finally, topics \#1 and \#2 seem generically positive (or possibly neutral) comments.
However, topic-word \textbf{Minecraft} stands out,
which may indicate attempts at translating some of Minecraft's
features as a \smm level;
the stronger association with easy levels
might indicate that such attempts tend to be uninteresting
from the perspective of platforming level design.

\section{Threats to Validity}
\label{sec:threats}

The operationalization of the concept of level \emph{difficulty}
is a fundamental threat to \emph{construct} validity.
While \vg*{difficulty}[clear rate] and \vg*{difficulty}[difficulty]
are reasonable proxies, they may fail to capture other
aspects of a level's difficulty, including a user's \emph{perception}; analyzing these aspects would require a different kind of study.

Since we analyzed observational data,
our statistical analyses merely report \emph{associations}
rather than genuine causal relations---which is a fundamental threat
to \emph{internal} validity.
This paper's replication package 
includes a structural analysis of causal relations,
which seems to be consistent with the main model presented in the paper;
while a reliable causal analysis requires controlled experiments,
even purely predictive models can give interesting insights.
\bertopic was trained on general text,
which may not adequately cover \smm's vernacular;
fine-tuning an NLP model on the texts commonly used in comments
may thus improve its effectiveness for our analysis.

Our NLP analysis was limited to English
descriptions and comments,
which may restrict the \emph{external} validity of our results
(i.e., their generalizability).
Extending our analysis to other linguistic groups
(e.g., Japanese, which is the second most used language in \smm)
belongs to future work.

\section{Related Work}
\label{sec:rw}

Difficulty is a fundamental characteristic
of videogame experience~\cite{ChenFlowGames2007},
which has been investigated in disparate ways~\cite{alexander2013investigation,aponte2009scaling,aponte2011difficulty}.
For \emph{platformers} specifically,
a very recent work
introduced a model to automatically evaluate their difficulty
based on a level's structure~\cite{Francillette2025};
applying this model to \smm levels is an interesting future work direction.
The \emph{Platformer Experience Dataset (PED)}~\cite{karpouzis2015a}
corpus collects various kinds of data
that capture the experience of Super Mario Bros.\ players
with rankings, game content recordings, and visual 
recordings of players.
Another approach~\cite{Rosa2021a} measures difficulty in platformers
through 
a machine learning model trained on both performance telemetry data
and emotional data estimated from electrodermal activity.
All these approaches focus on precisely measuring
the difficulty of individual levels and player runs.
In contrast, 
our paper mainly analyzes \emph{metadata},
such as a level's style or description;
by focusing on such data,
we have access to large amounts of user information,
including measures, such as the clear rate,
that are normally not available for single-player games.
Our approach is also justified by the specific game
we targeted:
in \smm, (some) players are also level designers,
which adds an interesting dimension to the analysis.
While metadata is arguably more coarse-grained than, say, a level's detailed structure,
our work demonstrated that it can provide complementary, broad-stroke insights.

A very different angle to analyze games
is as models of computation~\cite{ALOUPIS2015135,abel_et_al:LIPIcs.FUN.2024.1,jcai20:angrybirds}.
For example,
the undecidability~\cite{mithardnessgroup_et_al:LIPIcs.FUN.2024.22}
and the computational complexity~\cite{ALOUPIS2015135,demaine_et_al:LIPIcs.FUN.2016.13}
of classic Nintendo games,
in particular Mario games,
has been painstakingly studied.
NLP sentiment analysis has been applied to artifacts related to videogames,
including reviews~\cite{ViggiatoIEEE22}, chats~\cite{THOMPSON2017149}, and
player speech~\cite{sykownik2019can}. Other work 
focused on specific (sub)genres, like Virtual Reality~\cite{Guzsvinecz24}
or ``Souls-like'' games~\cite{Sicheng2024a};
their findings include that player negative emotions 
can also be influenced by cultural elements.

\section{Conclusions}
\label{sec:conclusion}

This paper analyzed a treasure trove of data
about \smm levels and users,
with the goal of studying
the factors that are associated with
a level's difficulty.
The main findings include:
\ifextended
\begin{enumerate}
\else
\begin{enumerate*}
\fi
\ifextended
\item Unsurprisingly, levels with long world records
  or that took a lot of attempts to upload
  are also harder for general players.
\fi
\item Levels that generate a lot of plays tend to be harder.
\item While most variables characterizing makers have negligible associations
  with level difficulty,
  makers with more experience tend to design easier levels
  (possibly because they are better designers).
\item Usually, older game styles (e.g., Super Mario Bros.~1)
  tend to be associated with easier levels than newer game styles
  (e.g., Super Mario World)---especially in popular levels.
\ifextended
\item Several associations tend to become stronger
  if we only consider popular levels.
\fi
\item Descriptions of the harder levels
  often refer to challenging game mechanics,
  whereas easy levels' descriptions may be bland or generic.
\item Comments expressing negative sentiments
  are more common in harder levels.
\item However, several frequently occurring comment topics in the same levels
  express appreciation---when the difficulty is justified by a high-quality design.
\ifextended
\item In contrast, easy levels
  tend to provoke comments that are often generically positive,
  or complain about an abuse of (visual) effects in level design.
\fi
\ifextended
\end{enumerate}
\else
\end{enumerate*}
\fi

%% \bibliographystyle{plain}

%% \bibliography{smm2}

\end{document}